\documentclass[aps, prl, superscriptaddress, twocolumn, floatfix]{revtex4-2}
\usepackage{graphicx, epsfig, color}
\usepackage{amssymb}
\usepackage{amsmath}
\usepackage{hyperref}
\usepackage{bm}
\usepackage{cleveref}
\begin{document}
\title{Multi-Particle Quantum Walks in a Dipole-Conserving Bose-Hubbard Model}
\author{Sooshin Kim}\email{sooshinkim@postech.ac.kr}\altaffiliation[current address: ]{Basic Science Research Institute, Pohang University of Science and Technology (POSTECH), Pohang 37673, Korea}
\affiliation{Department of Physics, Harvard University, Cambridge, Massachusetts 02138, USA}
\author{Byungmin Kang}
\affiliation{Department of Physics, Massachusetts Institute of Technology, Cambridge, Massachusetts 02139, USA}
\author{Perrin Segura}
\author{Yanfei Li}
\affiliation{Department of Physics, Harvard University, Cambridge, Massachusetts 02138, USA}
\author{Ethan Lake}
\affiliation{Department of Physics, Massachusetts Institute of Technology, Cambridge, Massachusetts 02139, USA}
\affiliation{Department of Physics, University of California Berkeley, Berkeley, California 94720, USA}
\author{Brice Bakkali-Hassani}
\affiliation{Department of Physics, Harvard University, Cambridge, Massachusetts 02138, USA}
\affiliation{Laboratoire Kastler Brossel, Coll\`ege de France, CNRS, ENS-PSL University, Sorbonne Universit\'e, 11 Place Marcelin Berthelot, 75005 Paris, France}
\author{Markus Greiner}
\affiliation{Department of Physics, Harvard University, Cambridge, Massachusetts 02138, USA}

\begin{abstract}
When particles move through a crystal or optical lattice, their motion can sometimes become frozen by strong external forces -- yet collective motion may still emerge through subtle many-body effects. In this work, we explore such constrained dynamics by realizing a dipole-conserving Bose-Hubbard model, where single atoms are immobile but pairs of particles can move cooperatively while preserving the system's center of mass, i.e. the overall dipole moment of the particle distribution. Starting from a one-dimensional chain of ultracold bosonic atoms in an optical lattice, we generate localized dipole excitations consisting of a hole and a doublon using site-resolved optical potentials and characterize their quantum walks and scattering dynamics. Our study provides a bottom-up investigation of a Hamiltonian with kinetic constraints, and paves the way for exploring low-energy phases of fractonic matter in existing experimental platforms.
\end{abstract}

\maketitle

\begin{figure}
	\centering
	\includegraphics{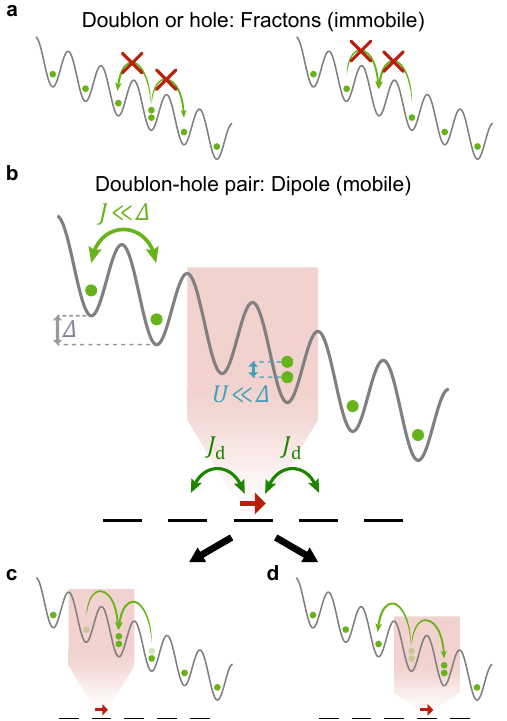}
	\caption{\textbf{Elementary excitations of a dipole-conserving Bose-Hubbard model at unity filling.} When the potential gradient $\Delta$ is much larger than the tunneling amplitude $J$ and on-site interaction $U$, the dynamics induced by the tilted Bose-Hubbard model $H_0$~\eqref{eq:tBH} effectively conserves the system's center of mass, equivalently described as the dipole moment of the particle distribution. \textbf{a,} While a doublon or a hole -- each representing an isolated fracton excitation -- cannot move individually along the chain, \textbf{b,} a dipole consisting of a hole-doublon pair can hop \textbf{c,} to the left or \textbf{d,} to the right. This is possible through second-order tunneling processes that preserve the total dipole moment. Here, the dipole is represented as a red arrow on a link between two adjacent sites, pointing in the direction of the dipole moment.}
	\label{Figure1}
\end{figure}

In many-body systems, limiting the mobility of individual particles can fundamentally alter the nature of collective dynamics. A well-known instance is the partially-filled lowest Landau level, where the kinetic energy is quenched into a highly degenerate flatband dominated by electron-electron interactions, and the resulting nonequilibrium dynamics can exhibit slow, subdiffusive relaxation due to emergent conservation laws~\cite{PhysRevLett.55.2095, PhysRevB.31.5280, PhysRevLett.95.266405, PhysRevB.102.195150, PRXQuantum.6.020321}. A simpler setting is provided by a one-dimensional (1D) lattice system with a strong linear potential [Fig.~\ref{Figure1}], where the energy is approximately set by the system's center of mass, which is identical to the dipole moment of the particle distribution. In this regime, energy conservation effectively suppresses individual particle motion and leads to the conservation of the dipole moment, a phenomenon known as Wannier-Stark localization~\cite{1962Wannier}. However, though single-particle mobility is suppressed, interactions still enable many-body dynamics through correlated hopping processes that preserve the dipole moment~\cite{2002SachdevGirvin}.

The constrained dynamics arising in the presence of a strong potential gradient belongs to a larger class of fracton models characterized by the conservation of multipole moments of a global charge, in this case the dipole moment associated with particle number~\cite{2018Pretko, 2019Nandkishore, 2020Pretko}. Fractons are excitations whose individual mobility is limited but which exhibit non-trivial dynamics when forming bound states [Fig.~\ref{Figure1}a,b]. Initially studied in the context of glassy spin dynamics and elasticity theory~\cite{2024Radzihovsky, 2005Chamon}, fractons have recently attracted interest in the study of out-of-equilibrium many-body dynamics, following predictions of phenomena such as Stark many-body localization~\cite{2019Pollmann_StarkMBL, 2021Monroe, 2021Wang}, anomalous particle diffusion~\cite{2020Nandkishore, 2020FeldmeierSalaPollmannKnap, 2020MorningstarKhemaniHuse, 2020Zhang, 2020Bakr}, and slow correlation spreading~\cite{2021FeldmeierKnap}. Multipole moment conservation often leads to Hilbert space fragmentation (HSF), with dynamics confined to many disconnected sectors beyond symmetry constraints~\cite{2019Pollmann_FragmentationAndLocalization, 2020KnapPollmann_Fragmentation, 2020GorshkovIadecola, khemani2020localization, 2022Regnault_Review, 2023Pollmann, 2024Nandkishore, honda_observation_2025}. Recent experiments in tilted 1D and 2D optical lattices have revealed HSF-induced non-ergodicity and fractonic excitations~\cite{2021BlochAidelsburger, 2023BlochAidelsburger, 2024PollmannBlochZeiher}. However, direct experimental observation of individual bound states formed by fractonic excitations has not been reported so far.

In this work, we generate and characterize mobile bound states of fractons using ultracold atoms in a strongly tilted Bose-Hubbard chain. Specifically, the single-site resolution of our quantum gas microscope allows us to prepare a localized hole-doublon pair on top of a unity-filled chain, and to track the evolution of this so-called dipole excitation using full-counting statistics. Additionally, we study the scattering dynamics between dipole excitations and reveal their strongly-interacting nature. Constrained-bosons dynamics in a one-dimensional lattice is theoretically expected to give rise to a variety of low-energy phases, including Mott insulating, dipole superfluid (Luttinger liquid), and dipole supersolid phases~\cite{sengupta_phases_2022, 2022Lake, 2023AltmanKnapFeldmeier, 2023Lake_Non-Fermi, 2023Lake_DC}. Our experiments provide experimental evidence that localized excitations can be used to probe these fractonic phases of matter~\cite{boesl_deconfinement_2024}.

\begin{figure*}
	\centering
	\includegraphics{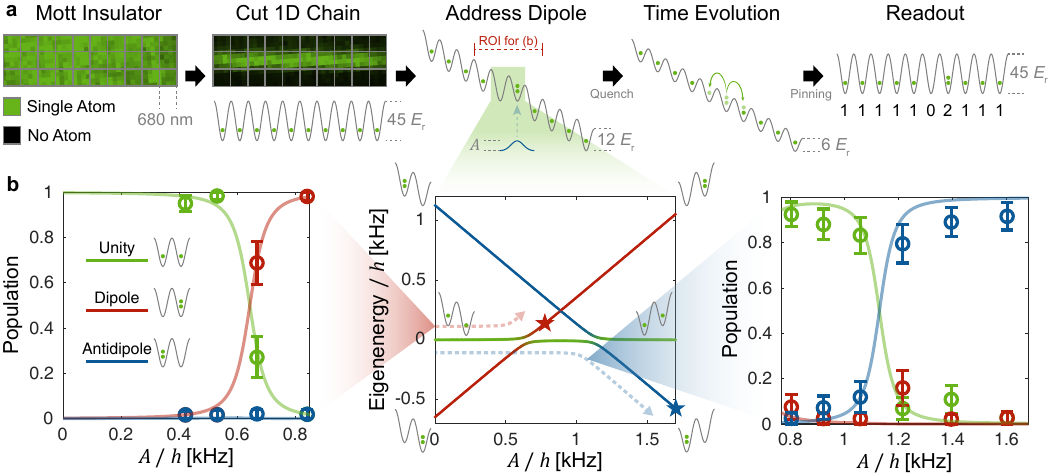}
	\caption{\textbf{Experimental protocol and preparation of a dipole state.} \textbf{a,} We isolate a 1D chain of atoms out of a unity-filled Mott insulator in a deep optical lattice. In the presence of a linear potential, a localized optical potential $A$ is then applied onto a single site to create a hole-doublon pair. Next, nonequilibrium dynamics is initiated in a shallow lattice by abruptly switching off $A$. After a variable evolution time, we project the many-body state onto the Fock basis by counting the number of atoms at each site. \textbf{b,} The center panel shows the energy spectrum for a two-particle, two-site system as a function of $A$. To create a dipole configuration (0, 2) from the initial state (1, 1), we adiabatically ramp $A$ from 0 along the red arrow. Alternatively, we create an antidipole configuration (2, 0) by suddenly increasing $A$ across the first avoided crossing and then adiabatically ramping $A$ along the blue arrow. On the left and right sides, we show the populations of the three configurations for different final values of $A$, measured over a 4-site region of interest (ROI) as indicated in the third illustration of panel a. Solid lines display numerical predictions for the populations assuming perfect adiabaticity across the first (left) and last (right) avoided crossings, respectively. Error bars denote 1$\sigma$ statistical uncertainties.}
	\label{Figure2}
\end{figure*}

We initiate our experiments by isolating a single chain from a unity-filled two-dimensional Mott insulator of ultracold $^{87}$Rb atoms using a high-resolution imaging system and site-resolved optical potentials, as shown in Fig.~\ref{Figure2}a~\cite{SM}. We restrict the analysis to a fixed region of interest (ROI) of ten sites, chosen within the larger Mott insulator. A linear potential with offset energy $\Delta = h \times 886(4)\,\text{Hz}$ per lattice site, where $h$ is Planck's constant, is then applied along the chain. The corresponding tilted Bose-Hubbard system is described by the Hamiltonian
\begin{align}
	\label{eq:tBH} 
	H_0 = & - J \sum_j \left(b_j b_{j + 1}^\dagger + \textrm{h.c.} \right) \nonumber \\
	&  + \frac{U}{2} \sum_j n_j (n_j - 1) - \Delta \sum_j n_j j,
\end{align}
where $b_j^\dagger$ ($b_j$) is the bosonic creation (annihilation) operator at site $j$, $n_j = b_j^\dagger b_j$ is the number operator, $J$ is the tunneling amplitude (which is negligible initially), and $U$ is the on-site interaction energy. To initialize the chain with a localized dipole excitation as shown in the third illustration of Fig.~\ref{Figure2}a, we start by coupling adjacent lattice sites with tunneling amplitude $J = h \times 21.2(5)\,\text{Hz}$. This is achieved by ramping down the lattice depth from $45 E_{\rm r}$ to $12 E_{\rm r}$ over 1~ms, where $E_{\rm r} = h \times 1.24\,\text{kHz}$ is the recoil energy. Note that many-body dynamics is still frozen at this stage~\cite{SM}. We then apply a localized optical potential to either pull a particle toward a lower-energy site (i.e. to the right), thereby creating a (0, 2) Fock configuration, which we refer to as a dipole excitation, or to push a particle toward a higher-energy site (i.e. to the left), forming a (2, 0) antidipole configuration.

Our initialization protocol is based on the adiabatic transfer of a single atom into an already occupied site, enabling the controlled preparation of doublons in a strongly tilted optical lattice, and is best understood from a minimal two-site model. Fig.~\ref{Figure2}b shows the corresponding energy spectrum in the case of two particles. The difference of potential energy between the right and left sites is given by $A - \Delta$, where $A$ results from the additional optical potential~\cite{SM}. In the absence of tunneling ($J = 0$), the eigenenergies corresponding to the (0, 2), (1, 1), and (2, 0) Fock configurations depend linearly on $A$. The presence of nonzero $J$ leads to three avoided crossings at $A = \Delta - U$, $\Delta$, and $\Delta + U$, where tunneling couples configurations with the same total potential and interaction energy. Our system is initially prepared with one particle per site, i.e. in the (1, 1) configuration. By adiabatically increasing $A$ from $0$ across the first avoided crossing over 125~ms [red arrow in Fig.~\ref{Figure2}b, center panel], the system is brought into the (0, 2) configuration, generating a dipole with a fidelity of 98(2)\% [Fig.~\ref{Figure2}b, left panel] after postselecting configurations containing four atoms within the ROI indicated in the center panel of Fig.~\ref{Figure2}a. To generate an antidipole, an adiabatic passage across the two consecutive avoided crossings results in a significantly lower fidelity. Indeed, the second energy gap  is an order of magnitude smaller than the first one because the (2, 0) and (0, 2) configurations are coupled only through second-order tunneling processes. Instead, we employ a two-step protocol: the first avoided crossing is passed diabatically in 1~ms, followed by an adiabatic passage of the third crossing in 100~ms [blue arrow in Fig.~\ref{Figure2}b, center panel], yielding an overall fidelity of 92(6)\% [Fig.~\ref{Figure2}b, right panel].

After creating the dipole or the antidipole, the lattice depth is reduced to $6 E_{\rm r}$ in 1~ms, yielding on-site interaction $U = h \times 203(4)\,\text{Hz}$ and tunneling amplitude $J = h \times 81(2)\,\text{Hz}$~\cite{SM, 2022Greiner}. The offset potential $A$ is then suddenly switched off, triggering the dynamics. In the regime $J, U \ll \Delta$, the total dipole moment $D \equiv \sum_j n_j j$ is conserved up to an exponentially-long prethermal timescale $\sim \frac{h}{J} \exp \big( \frac{\Delta}{J} \big) = 0.6(2)\,\text{s}$~\cite{khemani2020localization, boesl_deconfinement_2024}. In this regime of approximate dipole-moment conservation, a Schrieffer-Wolff transformation of the Hamiltonian $H_0$~\eqref{eq:tBH}, which effectively captures second-order dipole tunneling processes while suppressing single-particle motion, yields the following dipole-conserving Hubbard Hamiltonian:
\begin{align}
	\label{eq:dBH} 
		H = - \frac{J_{\rm d} }{2} \sum_j \left[ b_j^\dagger (b_{j + 1} )^2 b_{j + 2}^\dagger + \textrm{h.c.} \right] +  \frac{U_{\rm d} }{2} \sum_j n_j (n_j - 1).
\end{align}
In this expression, the effective dipole tunneling amplitude $J_{\rm d} = 2 \lambda^2 U = h \times 3.4(2)\,\text{Hz}$, where $\lambda = J / \Delta$ is the small parameter for the perturbative expansion, describes correlated hopping of two particles initially occupying the same site, as depicted in Fig.~\ref{Figure1}c,d, while $U_{\rm d} = (1 - 4 \lambda^2) U = h \times 196(4)\,\text{Hz}$ is the effective on-site interaction energy within the dipole-conserving subspace~\cite{SM}. The Hamiltonian~\eqref{eq:dBH} forbids single-particle tunneling and constitutes a Bose-Hubbard analog of the spin-1 system studied in the original works on HSF~\cite{2020KnapPollmann_Fragmentation, 2020KnapPollmann_Localization}. In particular, this model displays a phase transition from weak to strong fragmentation at the critical density of one boson per site~\cite{2023HuseSkinner}, the density realized in our experiment. The Schrieffer-Wolff transformation also produces a nearest-neighbor interaction term $V_{\rm d} \sum_j n_j n_{j+1}$ with strength $V_{\rm d} = 4 \lambda^2 U = h \times 6.8(5)\,\text{Hz}$, whose effects are discussed in the Supplements~\cite{SM}. The system evolves for a variable duration $t$ under this Hamiltonian, after which the lattice depth is increased to $45 E_{\rm r}$ in 1~ms to freeze the dynamics. Experiments are performed deep in the dipole--Mott-insulator (dipole-MI) regime, where spontaneous dipole-antidipole pair creation is suppressed and the number of dipole excitations is approximately conserved during the dynamics~\cite{2023AltmanKnapFeldmeier, 2023Lake_DC}. Full-counting statistics of the quantum state are obtained via fluorescence imaging, and only snapshots with the correct particle number ($N$) and dipole moment ($D$) are postselected to mitigate the influence of defects in the initial Mott insulator and infidelity in dipole preparation.

\begin{figure}
	\centering
	\includegraphics{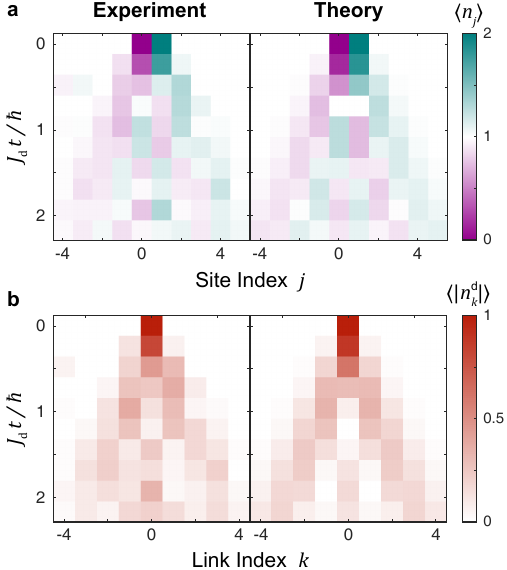}
	\caption{\textbf{Quantum walk of a single dipole.} Comparison between experimental data (left) and numerical simulations based on the dipole-conserving Hubbard model $H$~\eqref{eq:dBH} (right). Panels a and b show the atom-number and dipole-number densities, respectively~\cite{Jd}. The dipole wavepacket expands linearly in time while exhibiting density modulation indicative of quantum interference.}
	\label{Figure3}
\end{figure}

The evolution of the atomic density distribution as a function of time is shown in Fig.~\ref{Figure3}a. The density defects created by the doublon and hole spread across the chain, forming clear maxima and minima due to quantum interference in the coherent evolution. To highlight the correlated hole-doublon motion, in Fig.~\ref{Figure3}b, we re-express the data in terms of the local dipole charge $n_{k}^{\rm d}$, defined as
\begin{equation}
	\label{eq:Definition_Local_Dipole_Charge}
	n_{k}^{\rm d} \equiv - \sum_{j \le k} (n_j - \overline{n} ),
\end{equation}
which is associated with the link between sites $k$ and $k + 1$. Here, $\overline{n} = 1$ is the average filling number, so that the dipole charge naturally serves as a link degree of freedom~\cite{SM, 2023AltmanKnapFeldmeier}. With the hole (respectively doublon) initial position set to be $j = 0$ (respectively $j = 1$), the corresponding dipole is prepared on the link $k = 0$ between the two atomic sites. The first term of Eq.~\eqref{eq:dBH} describes the hopping of a dipole to a site on the right, akin to particle tunneling  described by $b_j b_{j+1}^\dagger$ in a Bose-Hubbard model. Therefore, the dynamics of a single dipole can be described as a quantum walk with tunneling amplitude $J_{\rm d}$, resulting in the density distribution
\begin{equation}
	\label{eq:QW}
	\langle | n_{k}^{\rm d} | \rangle_t = |\mathcal{J}_k (2J_{\rm d}t/\hbar)|^2,
\end{equation}
where $\mathcal{J}_k$ is the $k$-th Bessel function of the first kind and $\hbar = h / 2\pi$~\cite{SM, 2015Greiner,2004Mossmann}. Here, $| n_{k}^{\rm d} |$ denotes the magnitude of the local dipole charge, corresponding to the dipole density, since the sign of $n_{k}^{\rm d}$ merely distinguishes dipoles from antidipoles. A striking feature of Fig.~\ref{Figure3}b is the dipole density revival at $k = 0$ around two tunneling times, indicative of matterwave interference exhibited by the dipole. The coherent propagation of a single dipole reflects its stability as a quasiparticle in the dipole-MI regime. Note that in contrast, in the gapless dipole Luttinger-liquid regime, a local dipole excitation is expected to produce an asymmetric dipole density profile, characterized by diffusive waves at early times~\cite{boesl_deconfinement_2024}. 

\begin{figure}
	\centering
	\includegraphics{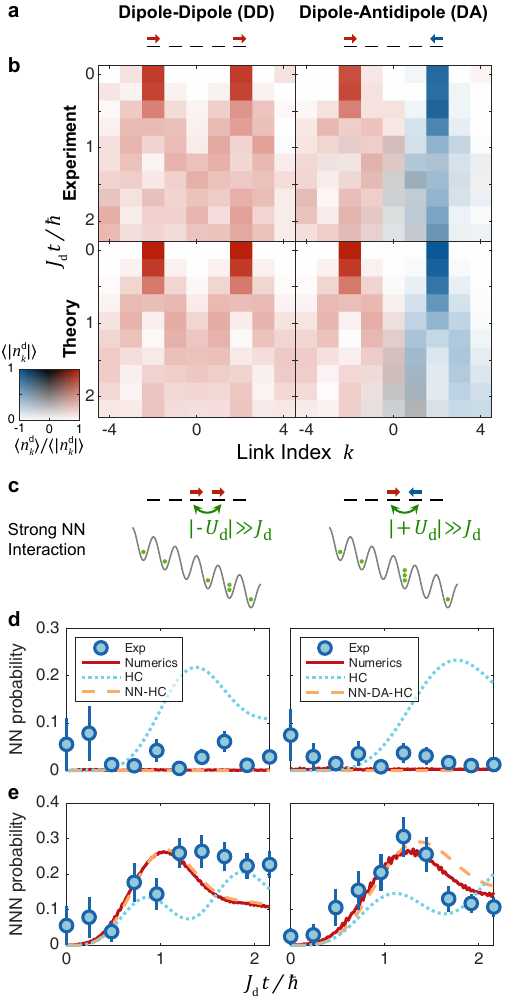}
	\caption{\textbf{Scattering dynamics of two dipoles.} \textbf{a,} We initialize a dipole-dipole (DD) or a dipole-antidipole (DA) pair with a four-link spacing. \textbf{b,} Density of dipole excitations and polarization $\langle n_k^{\rm d} \rangle / \langle |n_k^{\rm d}| \rangle$ at each link as a function of time~\cite{polarization}. \textbf{c,} Due to the energy cost $|\pm U_{\rm d}| \gg J_{\rm d}$ of occupying nearest-neighbor (NN) links, dipole excitations behave like hardcore (HC) NN particles. \textbf{d,} [resp. \textbf{e,}] Probability of finding a DD (left) or DA (right) pair in NN [resp. next-nearest-neighbor (NNN)] configurations, compared to the results of exact diagonalization simulation with the tilted Bose-Hubbard chain $H_0$ (red), as well as for the cases of two on-site HC bosons (blue) and two NN--HC bosons (yellow), with the same initial separation and tunneling amplitudes estimated for the dipole and antidipole~\cite{SM}. Error bars denote $1\sigma$ statistical uncertainties.}
	\label{Figure4}
\end{figure}

We now investigate the scattering dynamics of dipole-dipole (DD) and dipole-antidipole (DA) pairs. In our experiment, the two excitations are initially prepared four links apart [Fig.~\ref{Figure4}a]. We then monitor their correlated quantum walks, comparing with exact diagonalization simulations of the dynamics with no free parameter. Starting with the DD case, the left column of Fig.~\ref{Figure4}b shows the symmetric ballistic expansion of the two dipoles which begin to overlap for times $J_{\rm d} t / \hbar \gtrsim 1$, enabling us to probe their interactions via pair correlations extracted from full-counting statistics.

It is straightforward that a DD pair exhibits on-site hardcore (HC) interactions. Using Eq.~\eqref{eq:Definition_Local_Dipole_Charge}, configurations with two dipoles occupying the same link, $|\cdots, 0, 2, 0, \cdots\rangle_{\rm d}$, are forbidden because they map onto Fock states $|\cdots, 1, -1, 3, 1, \cdots\rangle$ which require negative density in the atom-number basis. In addition to this constraint, there exists an energy penalty $(-U_{\rm d})$ associated to nearest-neighbor (NN) configurations, $|\cdots, 0, 1, 1, 0, \cdots\rangle_{\rm d}$ because the corresponding atomic Fock states $|\cdots, 1, 0, 1, 2, 1, \cdots\rangle$ contain one less doublon excitation than the initial state [Fig.~\ref{Figure4}c]. Given that $U_{\rm d} \gg J_{\rm d}$, energy conservation leads to strong NN interactions between DD pairs described by the dipole-conserving Hamiltonian \eqref{eq:dBH} at $\overline{n} = 1$ in the dipole-MI regime, along with an on-site HC constraint.

To highlight these interactions, we plot the probabilities of finding configurations in which the dipole pair occupies NN or next-nearest-neighbor (NNN) links. In Fig.~\ref{Figure4}d, the NN configurations of the DD pair are suppressed at intermediate times, $1 \lesssim J_{\rm d} t / \hbar \lesssim 2$, compared to the dynamics of on-site--HC bosons having the same tunneling amplitude. Instead, our experimental and numerical results for the DD pair remain closer to that of our simulation for bosons with an NN--HC interaction. In addition, since NN configurations are energetically disfavored, the probability of finding the dipoles in the NNN configurations is relatively enhanced at intermediate times, as depicted in Figure~\ref{Figure4}e. These direct observations of suppressed NN and enhanced NNN coincidences indicate that, in our experimental regime, the scattering dynamics of the DD pair is governed by their strong NN interaction, effectively approaching the NN--HC limit.

Analogous to the DD case, a DA pair demonstrates a strong NN interaction. Having a dipole on the left NN link of an antidipole, $|\cdots, 0, 1, -1, 0, \cdots\rangle_{\rm d}$, which we term NN--DA configurations, requires the formation of a trion excitation of atoms, $|\cdots, 1, 0, 3, 0, 1, \cdots\rangle$, as illustrated in Fig.~\ref{Figure4}c. Hence, the dipole-conserving Hamiltonian $H$~\eqref{eq:dBH} imposes an energy penalty of $+U_{\rm d}$ on the NN--DA configurations, indicating a strong NN--DA interaction. The occurrence of the NN--DA and NNN--DA configurations are shown in Fig.~\ref{Figure4}d,e and compared to the numerically simulated dynamics of two distinguishable particles with either HC or NN--DA--HC interactions. The observed behavior agrees with the dynamics obtained for NN--DA--HC particles. Overall, these measurements establish that the DA pair is subject to an NN--HC constraint, as seen in the suppression of NN and enhancement of NNN configurations.

As a side remark, going beyond the dipole-conserving Hamiltonian $H$~\eqref{eq:dBH}, an antidipole is expected to tunnel more slowly than a dipole for non-zero $U/\Delta$. At leading order, the result of the Schrieffer-Wolff transformation for the tunneling amplitude $J_{\rm d}$ is modified by a relative correction $\pm U/\Delta$ for a dipole and an antidipole, respectively, with $U/\Delta = 0.22(1)$ in the present experiment~\cite{SM}. This asymmetry, which can be tuned and reduced by varying the tilt $\Delta$, is visible in the right column of Fig.~\ref{Figure4}b from the slower expansion rate of the antidipole at short times $J_{\rm d} t / \hbar \lesssim 1$. For the numerical simulations of the HC and NN--DA--HC particles in Fig.~\ref{Figure4}d,e, we thus employ asymmetric tunneling amplitudes matching those of the DA pair. Additionally, note that the deviation between experiment and numerics in Fig.~\ref{Figure4}b can be attributed to disorder of the lattice potential~\cite{SM}.

Expanding the scope, a notable feature of dipole-antidipole scattering is its inherent directionality. While Fock configurations with a dipole as the left nearest neighbor of an antidipole (NN--DA) are present, the opposite configurations, where an antidipole lies to the left of a dipole ($|\cdots, 0, -1, 1, 0, \cdots\rangle_{\rm d}$, denoted NN--AD), map onto unphysical Fock states $|\cdots, 1, 2, -1, 2, 1, \cdots\rangle$ and are excluded from the Hilbert space. This indicates that the NN--AD configurations are intrinsically hardcore, whereas the hardcore nature of NN--DA configurations relaxes at larger $J_{\rm d} / U_{\rm d}$. This constraint corresponds to a matched-parenthesis rule, an exotic structure also present in the highly-entangled Motzkin spin chain~\cite{SM, 2016Shor, 2017Movassagh}, and is expected to influence the dynamics once the system leaves the deep dipole-MI regime and approaches the dipole Luttinger-liquid phase with $\overline{n} = 1$. Finally, we note that the term $V_{\rm d}$ is expected to induce NNN interactions, which, however, remain beyond our current detection capability~\cite{SM}.

In summary, we have realized a kinetically constrained quantum system governed by a dipole-conserving Bose-Hubbard Hamiltonian, where single-particle motion is suppressed by a strong potential gradient and dipole moment conservation emerges as an effective symmetry. Using site-resolved optical potentials, we initialize localized hole-doublon (dipole) and doublon-hole (antidipole) excitations and track their nonequilibrium dynamics, which reveal strongly-correlated multi-particle quantum walks. We further demonstrate the strongly-interacting nature of dipole excitations through scattering experiments, where suppressed NN and enhanced NNN configurations provide clear evidence of strong NN interactions.

Looking forward, our platform enables exploration of ground-state phases predicted in the dipole-conserving Bose-Hubbard model and governed by the effective coupling parameter $J/\Delta$~\cite{2023Lake_DC, 2023AltmanKnapFeldmeier}, where increasing $J/\Delta$ (and thereby $J_{\rm d}/U_{\rm d}$) is expected to unveil rich phases such as fractured Bose droplets and dipole Luttinger liquids at higher fillings. In particular, the expansion dynamics of localized excitations, as realized here for dipole excitations, should provide a sensitive probe of these underlying phases~\cite{2024HanLee, boesl_deconfinement_2024}. More broadly, the matched-parenthesis constraint links our experiment to the Motzkin spin chain, where it underlies both Hilbert-space fragmentation and the emergence of volume-law entangled ground states in a local Hamiltonian~\cite{2016Shor, 2017Movassagh}, highlighting the richness of the effective dipole dynamics and motivating further exploration.

\begin{acknowledgments}
We acknowledge helpful discussions with T.~Chalopin, A.~Fabre, J.~Feldmeier, S.~Jang, Y.~Kim, R.~Vatr\'e, and M.~Xu. B.~K. acknowledges discussions with P.~A.~Lee. This work was supported by grants from the National Science Foundation, the Gordon and Betty Moore Foundations EPiQS Initiative, an Air Force Office of Scientific Research MURI program, and an Army Research Office MURI program. B.~K. is supported by DOE office of Basic Sciences Grant No. DE-FG02-03ER46076 (theory). E.~L. is supported by a Miller Research Fellowship.
\end{acknowledgments}

M.~G. is a cofounder and shareholder of QuEra Computing. All other authors declare no competing interests.

\bibliography{References}

\begin{thebibliography}{59}
\bibitem [{pol()}]{polarization}%
  \BibitemOpen
  \href@noop {} {\bibinfo {title} {{The local polarization reflects the average
  sign of dipole excitations on a given link, a feature not captured by the
  density of dipole excitations $\langle |n_k^{\rm d}| \rangle$, and is
  therefore essential for describing DA scattering.}}}\BibitemShut {Stop}%
\end{thebibliography}

\appendix

\clearpage

\section{SUPPLEMENTAL MATERIAL}

\subsection{Calibration of Bose-Hubbard parameters}

To calibrate the parameters of the tilted Bose-Hubbard model~\eqref{eq:tBH}, we induce quantum walks in a flat one-dimensional lattice, i.e. without applying a potential gradient, yielding tunneling amplitudes of $J = h \times 21.2(5)\,\text{Hz}$ and $81(2)\,\text{Hz}$ at lattice depths of $12\,E_{\rm r}$ and $6\,E_{\rm r}$, respectively, where $E_{\rm r} = h \times 1.24\,\text{kHz}$ is the recoil energy and $h$ Planck’s constant. Photon-assisted tunneling in a strongly tilted lattice provides the on-site interaction energy, $U = h \times 203(4)\,\text{Hz}$ at $6\,E_{\rm r}$, and the tilt per site, $\Delta = h \times 886(4)\,\text{Hz}$. Details of these calibration methods are given in Ref.~\cite{2019LukinGreiner}.

\subsection{Experimental sequence}

In our experiments, we first prepare a Bose-Einstein condensate of $^{87}$Rb atoms in the $|F = 1, m_F = -1 \rangle$ hyperfine state. The atoms are loaded into a single two-dimensional layer of a blue-detuned square optical lattice with lattice constant $a = 680\,\text{nm}$ and depth $45\,E_{\rm r}$, producing a Mott insulator~\cite{2009Greiner}. Next, we apply an optical potential shaped by a digital micromirror device (DMD) and projected through our microscope objective to isolate a one-dimensional chain of atoms from the $n = 1$ region of the Mott insulator with high fidelity~\cite{2015Greiner}. The chain length, typically over ten sites, is determined by the initial size of the Mott insulator.

To generate a dipole excitation, we apply a linear magnetic field gradient that provides a tilted potential $\Delta = h \times 886(4)\,\text{Hz}$ per lattice site along the chain. The lattice depth is subsequently lowered to $12\,E_{\rm r}$ within $1\,\text{ms}$, enabling tunneling with amplitude $J = h \times 21.2(5)\,\text{Hz}$. In this regime, the dynamics remain frozen since the effective dipolar tunneling is only $0.28(1)\,\text{Hz}$. A Gaussian potential with RMS width of $0.25$ sites, created by a second DMD, is then superimposed at the center of a lattice site to provide an on-site offset energy $A$. As described in the main text, $A$ is gradually increased from 0 to a finite value to realize either a dipole or an antidipole. A single linear ramp generates a dipole, while two piecewise linear ramps are used to form an antidipole.

\begin{figure*}
	\centering
	\includegraphics{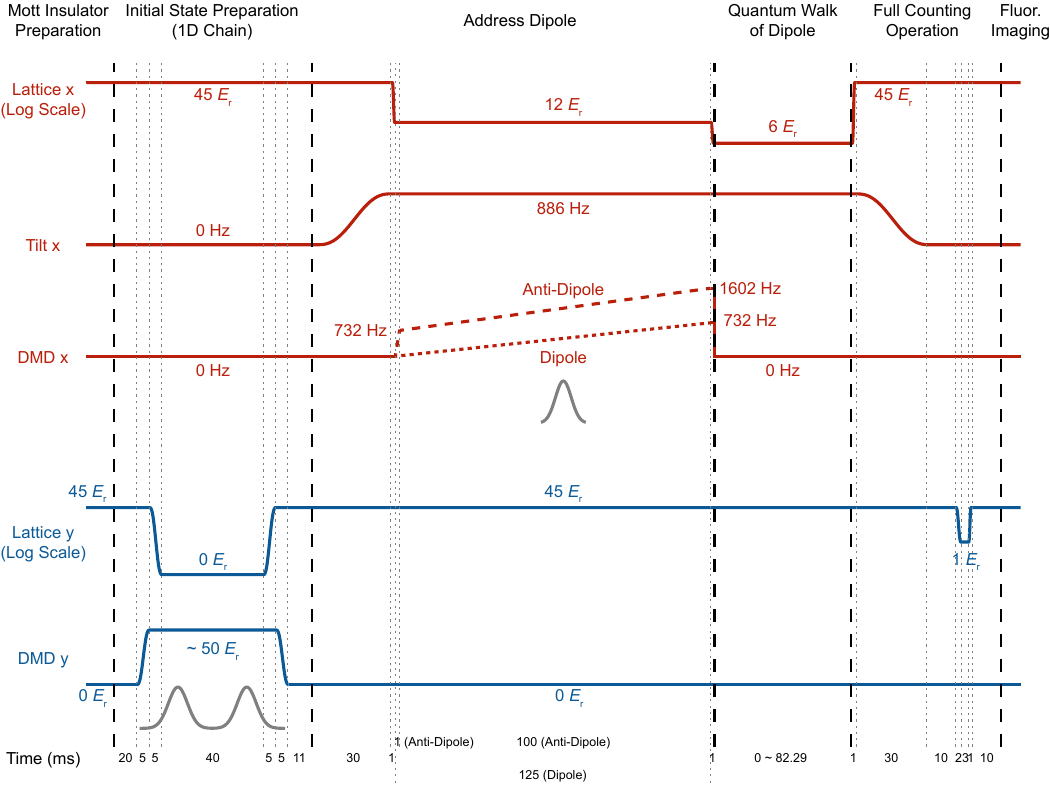}
	\caption{\textbf{Experimental sequence.} Major experimental parameters are shown as functions of time. An optical potential shaped with a DMD is projected onto a Mott insulator to isolate a unity-filled one-dimensional chain. The chain is then tilted along its axis by adiabatically ramping up a magnetic field gradient. A Gaussian potential from a second DMD is subsequently applied to prepare either a dipole or an antidipole. This pinning potential is then abruptly switched off to initiate the nonequilibrium dynamics. Finally, the atoms are expanded orthogonally to the chain and imaged individually by fluorescence.}
	\label{FigureS_Sequence}
\end{figure*}

After state preparation, the lattice depth is ramped down to $6\,E_{\rm r}$ within $1\,\text{ms}$, followed by a sudden quench of $A$ back to zero, which initiates the dipole quantum walk. After a variable evolution time $t$, the dynamics are frozen by increasing the lattice depth back to $45\,E_{\rm r}$ within $1\,\text{ms}$. Finally, the atoms are expanded along the direction perpendicular to the chain and imaged by fluorescence to measure the occupation of each lattice site~\cite{2016KaufmanGreiner}. To gather statistics on these site-resolved snapshots, we repeat the sequence between 50 and 900~times depending on the quantum walk duration. The detailed ramp profiles of the relevant experimental parameters are shown in Fig.~\ref{FigureS_Sequence}.

\subsection{Numerics}

We perform exact-diagonalization (ED) calculations to provide theoretical expectations for all figures in the main text and Supplemental Materials. The time-dependent Schrödinger equation for the tilted Bose-Hubbard Hamiltonian $H_0(t)$, with parameters of Eq.~\eqref{eq:tBH} varied according to the lattice ramps following dipole excitation preparation (described in a later section), is numerically solved using a Krylov-subspace method with Trotterization~\cite{1998Sidje}. At each time step, Fock states with the wrong total dipole moment $D$ are projected out before extracting observables. To reach larger system sizes (up to 14 sites at unit filling), the Hilbert space is truncated in two ways. First, the on-site occupancy is restricted to a maximum of $n_{\max}=4$ atoms per site, justified by the large energy cost of higher occupations. Second, the dipole moment deviation from the initial state is limited by restricting single-particle motion to at most five successive hopping processes in one direction, corresponding to a maximum relative dipole moment change of $5/N$ for $N$ particles. We verified that these truncations do not affect our numerical predictions for system sizes up to 10 sites.

\subsection{Postselection}

\begin{figure}
	\centering
	\includegraphics[width = \columnwidth]{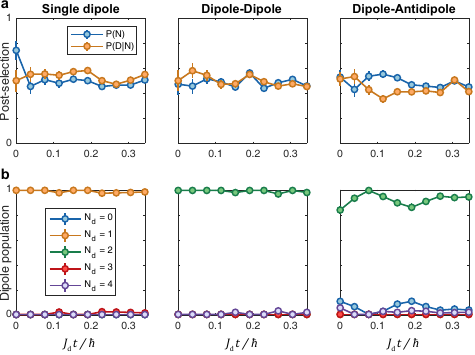}
	\caption{\textbf{Verification of postselection.} \textbf{a,} Postselection rate $P(N)$ for atom number conservation (blue dots), and $P(D|N)$ for dipole moment conservation conditioned on atom-number conservation (orange dots). \textbf{b,} Probability of total dipole excitation number with atom number and dipole moment postselection.}
	\label{FigureS_Postselection}
\end{figure}

In the experimental data reported in the main text, we apply two levels of postselection. First, we retain only snapshots with the correct total atom number within a region of interest of ten sites (ROI). This step eliminates errors due to imperfect initial Mott insulator preparation, atom loss during the initialization of dipole excitations and due to heating in subsequent steps. For quantum-walk data at finite evolution time, this number-conserving postselection rate lies between 40 and 60\% for a ROI of 10~sites with the initial dipole(s) centered. The same procedure is used in Fig.~\ref{Figure2}b when estimating the fidelities of dipole and antidipole generation within a ROI of 4~sites. To remove errors in the dipole-generation process, we further postselect snapshots with the desired total dipole moment within the number-conserving shots. The resulting dipole-conserving postselection rates for single dipoles, dipole–dipole pairs, and dipole–antidipole pairs are shown in Fig.~\ref{FigureS_Postselection}a. Importantly, Fig.~\ref{FigureS_Postselection}b also confirms the approximate conservation of the number of dipole excitations during the dynamics, consistent with the discussion in the main text. 

\subsection{Lattice ramp}

\begin{figure}
	\centering
	\includegraphics{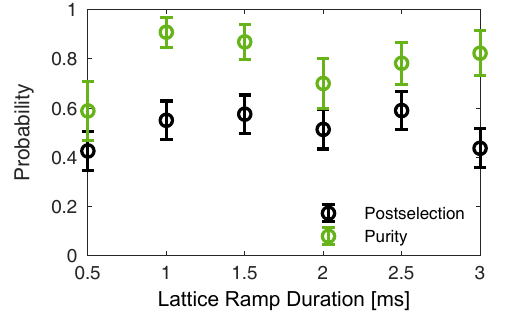}
	\caption{\textbf{Scan for optimal lattice ramp duration.} Postselection rate for total atom number conservation over 10~sites and purity with respect to the initial Fock state are shown. A strongly tilted ($\Delta = h \times 997(4)\,\text{Hz}$) unity-filled chain is prepared, the lattice depth is reduced from $45\,E_{\rm r}$ to $6\,E_{\rm r}$ for variable durations, held for 10~ms, and then ramped back with the same duration.}
	\label{FigureS_Lattice_Ramp}
\end{figure}

Two parameters are ramped in the experimental sequence: the lattice depth and the site-resolved optical potential $A$ (discussed in the main text). The lattice ramp must balance two competing energy scales. A ramp that is too fast compared to the tilt $\Delta$ induces excitations outside the effective dipole-conserving subspace. A ramp that is too slow compared to the effective dipole tunneling $J_{\rm d}$ allows dipole dynamics to occur already during the ramp, so that the prepared excitation is no longer well defined. The suitable regime is therefore one where the ramp is adiabatic with respect to $\Delta$ but diabatic with respect to $J_{\rm d}$~\cite{2022Greiner}. We tested different ramp durations by preparing a unity-filled one-dimensional chain in a deep $45\,E_{\rm r}$ lattice with a strong tilt $\Delta = h \times 997(4)\,\text{Hz}$. The depth was lowered to $6\,E_{\rm r}$ over a variable time, held for 10~ms, and then ramped back to $45\,E_{\rm r}$ in the same duration. We measured the purity of the initial unity-filled Fock state and the number-conserving postselection rate within a 10-site ROI. As shown in Fig.~\ref{FigureS_Lattice_Ramp}, ramps longer than 1~ms preserve the dipole moment best, and we therefore use a 1~ms ramp for all experiments.

\subsection{$\Delta$ window for dipole-conserving dynamics}

\begin{figure}
	\centering
	\includegraphics{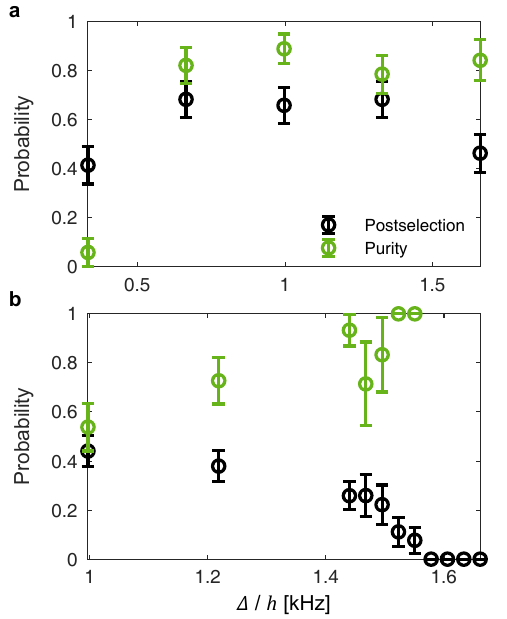}
	\caption{\textbf{Tilt window for dipole-conserving dynamics.} Postselection rate and purity are measured after lowering the lattice depth from $45\,E_{\rm r}$ to $6\,E_{\rm r}$, holding for \textbf{a,} 0~ms or \textbf{b,} 200~ms, and ramping back to $45\,E_{\rm r}$. At $\Delta/h \approx 1\,\text{kHz}$ the system can be mapped to the dipole-conserving model with high fidelity, while for larger $\Delta$ atom losses increase due to resonances with higher bands.}
	\label{FigureS_Tilt}
\end{figure}

The range of tilt $\Delta$ that supports dipole-conserving dynamics is limited. For small $\Delta$, single-particle tunneling is no longer suppressed and dipole conservation breaks down. For large $\Delta$ at finite lattice depth, resonances with higher Bloch bands enable long-range tunneling processes across several sites and lead to atom loss~\cite{1999Korsch, 2008Wimberger}. Figures~\ref{FigureS_Tilt}a,b show the corresponding measurements: for $\Delta/h \gtrsim 1.5\,\text{kHz}$, losses increase due to higher-band resonances, while after a 200~ms hold the purity also decreases near $\Delta/h \approx 1\,\text{kHz}$, consistent with resonances to higher bands.

\subsection{Influence of potential disorder on scattering dynamics}

\begin{figure}
	\centering
	\includegraphics[width = \columnwidth]{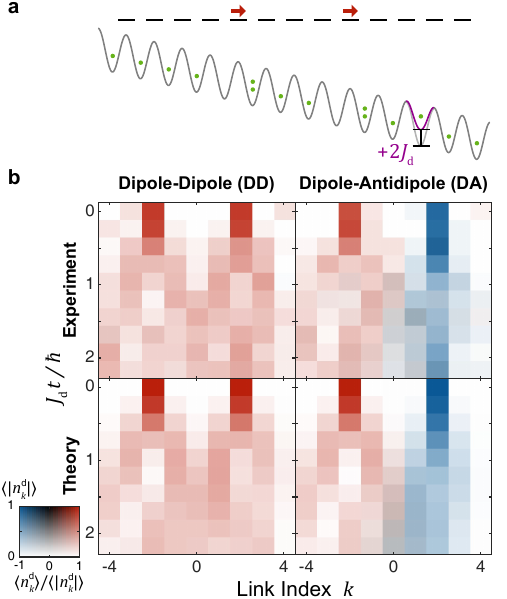}
	\caption{\textbf{Influence of potential disorder on the scattering dynamics of two dipoles.} \textbf{a,} Initialization of a dipole–dipole (DD) pair with an additional disorder potential $V=+2J_{\rm d}$ applied three sites to the right of the pair. \textbf{b,} Time evolution of the dipole density $\langle |n_k^{\rm d}| \rangle$ and polarization $\langle n_k^{\rm d} \rangle / \langle |n_k^{\rm d}| \rangle$.}
	\label{FigureS_Disorder}
\end{figure}

In Fig.~\ref{Figure4} of the main text, deviations between experiment and numerics can be accounted for by the presence of potential disorder along the chain. To test this, we performed numerical simulations of the tilted Bose–Hubbard model $H_0$ on system sizes up to $L=14$ with additional on-site potential offsets. Adding a local disorder potential $V = 2J_{\rm d} = 8.2\,\text{Hz}$ qualitatively reproduces the density profiles observed in Fig.~\ref{Figure4}b, as illustrated in Fig.~\ref{FigureS_Disorder}. This level of disorder is consistent with the intrinsic potential variations of our optical lattice.

\subsection{Derivation of the dipolar Bose-Hubbard model}

In this section, we derive the effective Hamiltonian for the tilted Bose-Hubbard model in the strong tilting limit, following the approach employed in, e.g., Refs.~\onlinecite{2021BlochAidelsburger, 2023Lake_DC}. To this end, we employ the Schrieffer-Wolff transformation to derive the effective dipole Bose-Hubbard model.  We begin with the following tilted Bose-Hubbard Hamiltonian:
\begin{align}
	{H}_0 = &-\Delta \sum_j n_j j + \frac{U}{2} \sum_j n_j (n_j - 1) \nonumber \\
	&- J \sum_{j} \big( b_j b_{j+1}^\dagger + b_j^\dagger b_{j + 1} \big) \nonumber \\
	\equiv &-\Delta \Big( {D} - \lambda({V} + {T} ) \Big) ,
\end{align}
where we introduced ${D} = \sum_j n_j j$, ${V} = \frac{U}{2J} \sum_j n_j (n_j - 1)$, ${T} = - \sum_j \big( b_j b_{j+1}^\dagger + b_j^\dagger b_{j + 1} \big)$, and $\lambda = J/\Delta$. We are interested in a regime where $|\lambda| \ll 1$.

Note that $\lambda {T}$ is off-diagonal with respect to ${D}$. We therefore apply the Schrieffer-Wolff transformation to perturbatively get rid of off-diagonal terms. Let us construct an anti-Hermitian operator ${S}$ where $e^{{S} } H_0 e^{-{S} }$ commutes with ${D}$. Such ${S}$ can be constructed perturbatively in $\lambda$. So, we set ${S} = \lambda {S}_1 + \lambda^2 {S}_2 + \lambda^3 {S}_3 + \ldots$, where each ${S}_l$ is independent of $\lambda$. The effective Hamiltonian ${H}_\textrm{eff}$ can be expressed as
\begin{widetext}
	\begin{align}
		{H}_{\rm eff} &= e^{{S}} {H}_0 e^{-{S}} = {H}_0 + \big[ {S}, {H}_0 \big] + \frac{1}{2} \big[ {S}, \big[ {S}, {H}_0 \big] \big] + \frac{1}{6} \big[ {S}, \big[ {S}, \big[ {S}, {H}_0 \big] \big] \big] + \ldots \nonumber \\
		&= {D} + \lambda \Big(- {V} - {T} + \big[{S}_1, {D} \big] \Big) + \lambda^2 \bigg( -\big[ {S}_1, {V} + {T} \big] + \frac{1}{2} \big[ {S}_1, \big[ {S}_1, {D} \big] \big] + \big[ {S}_2, {D} \big] \bigg) \nonumber \\
		&\quad + \lambda^3 \bigg( -\frac{1}{2} \big[ {S}_1, \big[ {S}_1, {V} + {T} \big] \big] - \big[ {S}_2, {V} + {T} \big] + \frac{1}{6} \big[ {S}_1, \big[ {S}_1, \big[ {S}_1, {D} \big] \big] \big] + \frac{1}{2} \big[ {S}_1, \big[ {S}_2, {D} \big] \big] \nonumber \\ 
		&\qquad \qquad + \frac{1}{2} \big[ {S}_2, \big[ {S}_1, {D} \big] \big] + \big[ {S}_3, {D} \big] \bigg) + \mathcal{O} \big( \lambda^4 \big) .
	\end{align}
	We choose ${S}_l$ in such a way that it cancels the off-diagonal terms order-by-order. This can be done as follows. We choose ${S}_1$ such that 
	\begin{equation}
		\label{eq:S1-D}
		[{S}_1, {D}] = {T}
	\end{equation} 
	holds. Then the off-diagonal terms appearing in $\mathcal{O}(\lambda)$ cancel and the effective Hamiltonian becomes
	\begin{align}
		{H}_\textrm{eff} = & {D} - \lambda {V} + \lambda^2 \Big( - \frac{1}{2} [ {S}_1, {T} ] - [ {S}_1, {V}] + [ {S}_2 , {D}] \Big) \nonumber \\
		&+ \lambda^3 \Big( - \frac{1}{3} [{S}_1, [{S}_1, {T}]] - \frac{1}{2} [{S}_1, [{S}_1, {V}]] - \frac{1}{2} [{S}_2, {T}] - [{S}_2, {V}] + \frac{1}{2} [{S}_1, [{S}_2, {D}]] + [{S}_3, {D}] \Big) + \mathcal{O} ( \lambda^4) . 
	\end{align}
	We now choose ${S}_2$ in such a way that $[{S}_2, {D}]$ cancels the off-diagonal terms at $\mathcal{O} (\lambda^2)$. Note that while $[{S}_1, {V}]$ is off-diagonal, $[{S}_1, {T}]$ can contain diagonal terms in general. Moreover, since ${D}$ is diagonal, $[{S}_2, {D}]$ cannot contain diagonal terms. We therefore choose an off-diagonal ${S}_2$ such that 
	\begin{align}
		\label{eq:S2-D}
		[{S}_2, {D}] = [{S}_1, {V}] + \frac{1}{2} \Big( [{S}_1, {T}] - \mathcal{P} [{S}_1, {T}] \mathcal{P} \Big)
	\end{align}
	holds, where $\mathcal{P}$ is the projector onto ${D} = D_\textrm{tot}$ subspace with $D_\textrm{tot}$ being the total dipole moment. For example, if the initial state is a product state, $D_\textrm{tot}$ is the total dipole moment of the initial state and $\mathcal{P}$ is the projector onto the configurations having the total dipole moment $D_\textrm{tot}$. This choice of ${S}_2$ further simplifies the effective Hamiltonian as 
	\begin{align}
		{H}_\textrm{eff} = & {D} - \lambda {V} - \lambda^2 \frac{1}{2} \Big( \mathcal{P} [ {S}_1, {T} ] \mathcal{P} \Big) \nonumber \\
		&+ \lambda^3 \Big( - \frac{1}{12} [{S}_1, [{S}_1, {T}]] - \frac{1}{4} \big[ {S}_1, \mathcal{P} [{S}_1, {T}] \mathcal{P} \big] - \frac{1}{2} [{S}_2, {T}] - [{S}_2, {V}] + [{S}_3, {D}] \Big) + \mathcal{O} ( \lambda^4) . 
	\end{align}
	We choose ${S}_3$ so that $[{S}_3, {D}]$ cancels all the off-diagonal terms appearing at $\mathcal{O} (\lambda^3)$. Since $[{S}_2, {V}]$ and $[{S}_1, \mathcal{P} [{S}_1, {T}] \mathcal{P}]$ are off-diagonal terms while $[{S}_1, [{S}_1, {T}]]$ and $[{S}_2, {T}]$ can contain diagonal terms, we demand 
	\begin{equation}
		\label{eq:S3-D}
		[{S}_3, {D}] = \frac{1}{12} \Big( [{S}_1, [{S}_1, {T}]] - \mathcal{P} [{S}_1, [{S}_1, {T}]] \mathcal{P} \Big) + \frac{1}{4} \big[{S}_1, \mathcal{P} [{S}_1, {T}] \mathcal{P} \big] + \frac{1}{2} \Big( [{S}_2, {T}] - \mathcal{P} [{S}_2, {T}] \mathcal{P} \Big) + [{S}_2, {V}] ,
	\end{equation}
	which simplifies the effective Hamiltonian as 
	\begin{equation}
		\label{eq:H-eff-lambda-3}
		{H} =  {D} - \lambda {V} - \lambda^2 \frac{1}{2} \Big( \mathcal{P} [ {S}_1, {T} ] \mathcal{P} \Big) -\lambda^3 \Big( \frac{1}{12} \mathcal{P} [{S}_1, [{S}_1, {T}]] \mathcal{P} + \frac{1}{2} \mathcal{P} [{S}_2, {T}] \mathcal{P} \Big) + \mathcal{O} ( \lambda^4) . 
	\end{equation}
	
	To get the final expression for ${H}_\textrm{eff}$ Eq.~\eqref{eq:H-eff-lambda-3}, we need to solve a series of equations Eqs.~\eqref{eq:S1-D},~\eqref{eq:S2-D}, and~\eqref{eq:S3-D} for ${S}_1$, ${S}_2$, and ${S}_3$. At first sight, this sounds like a complicated task since the equations involve many commutators. However, there is a simple way to find the solutions, which we now explain. Suppose we apply the operator $\sum_j b_{j+1}^\dagger b_j$ on a state, where the state has a well-defined total dipole moment $D_\textrm{tot}$. Then after acting the operator on the state, the resulting state has the total dipole moment $D_\textrm{tot} + 1$ due to one-site hopping terms. This fact can be recast as $[{D}, \sum_j b_{j+1}^\dagger b_j] = (+1) \sum_j b_{j+1}^\dagger b_j$. Similarly, we get $[{D}, \sum_j b_j^\dagger b_{j+1}] = (-1) \sum_j b_j^\dagger b_{j+1}$. In general, for a given operator ${O}$, we write ${O}$ as the sum 
	\begin{equation}
		{O} = \sum_{m \in \mathbb{Z}} {O}^{(m)} , \quad \textrm{where } [ {D}, {O}^{(m)}] = m {O}^{(m)}. 
	\end{equation}
	Then for an off-diaogonal operator ${O}$, (i.e., ${O}^{(0)} = 0$), the solution to $[{S}, {D}] = {O}$ is nothing but ${S} = - \sum_{m \in \mathbb{Z} \setminus \{0\}} \frac{1}{m} {O}^{(m)}$ : 
	\begin{equation}
		[{S}, {D}] = - \sum_{m \in \mathbb{Z} \setminus \{0\}} \frac{1}{m} [{O}^{(m)}, {D}] = \sum_{m \in \mathbb{Z} \setminus \{ 0 \}} {O}^{(m)} = {O}.
	\end{equation}
	Therefore, we set 
	\begin{equation}
		{S}_1 = \sum_j \big( b_{j+1}^\dagger b_j - b_j^\dagger b_{j+1} \big) , 
	\end{equation}
	which satisfies Eq.~\eqref{eq:S1-D}. After some tedious calculations, we get 
	\begin{align}
		[{S}_1, {T}] &= 0 \nonumber \\
		[{S}_1, {V}] &= \frac{U}{J} \sum_j \Big( n_j (b_{j+1}^\dagger b_j - b_{j-1}^\dagger b_j ) + (b_j^\dagger b_{j+1} - b_j^\dagger b_{j-1} ) n_j \Big) . 
	\end{align}
	Then the solution to Eq.~\eqref{eq:S2-D} is given by 
	\begin{equation}
		{S}_2 = \frac{U}{J} \sum_j \Big( - n_j (b_{j+1}^\dagger b_j + b_{j-1}^\dagger b_j ) + (b_j^\dagger b_{j+1} + b_j^\dagger b_{j-1} ) n_j \Big) . 
	\end{equation}
	After another lengthy calculation, we get 
	\begin{equation}
		\mathcal{P} [ {S}_2, {T} ] \mathcal{P} = - \frac{2U}{J} \sum_j \big( b_{j+1}^\dagger (b_j)^2 b_{j-1} + b_{j+1} (b_j^\dagger)^2 b_{j-1} \big) + \frac{8U}{J} \sum_j n_{j+1} n_j - \frac{4U}{J} \sum_j n_j (n_j - 1).
	\end{equation}
	Putting back the overall $(-\Delta)$ factor, the effective Hamiltonian Eq.~\eqref{eq:H-eff-lambda-3} is given by the following dipolar Bose-Hubbard model presented in the main text: 
	\begin{align}
		{H}_\textrm{dBH} = &- \Delta \sum_j j n_j - U \lambda^2 \sum_j \big( b_{j+1}^\dagger (b_j)^2 b_{j-1}^\dagger + b_{j+1} (b_j^\dagger)^2 b_{j-1} \big) + \frac{U}{2} (1 - 4 \lambda^2 ) \sum_j n_j (n_j - 1) + 4 U \lambda^2 \sum_j n_{j+1} n_j \nonumber \\
		&+ \mathcal{O} \big( U \lambda^3, J \lambda^3 \big) , 
	\end{align}
\end{widetext}
where $\lambda = J/\Delta$. This result differs from the expression of Ref.~\cite{2023Lake_DC} by a factor of $2$ in the coefficient of the $n_j n_{j+1}$ term, the discrepancy being caused by a typo in that reference.

\subsection{Dipole operator}

 In this section, we discuss how one can further simplify the dipolar Bose-Hubbard model by deriving the effective Hamiltonian in terms of dipole operators, which will turn out to be characterized by a biased hopping term and an effective next nearest neighbor interaction term. This effective Hamiltonian is most useful when the dipole density in the system is low, where it allows for a more succinct description of dipole quantum walks. 

There are two important facts about our experiment that we will employ in our construction of the effective Hamiltonian. First, the (renormalized) on-site Hubbard interaction is always much larger than the dipole hopping term strength and other parameters in the effective dipolar Hamiltonian. Second, our experiment always operates at an average boson density of $\overline{n} = 1$. These two facts prompt us to truncate the Hilbert space by restricting the boson occupation number $n_i \le 2$ at each site. This constraint together with dipole conservation produce rather unusual dynamics for the dipole excitations, in a way that we now explain.

To understand the effective dynamics in this truncated Hilbert space, let us consider how the dipole hopping term $T_\textrm{d} = -\frac{J_\textrm{d}}{2} \sum_j \big( b_{j-1} (b_{j}^\dagger)^2 b_{j+1} + b_{j-1}^\dagger (b_{j})^2 b_{j+1}^\dagger \big)$ acts in the restricted Hilbert space. In this case, the dipole hopping term on a collection of three neighboring sites becomes
\begin{align}
	\label{eq:dipole-hopping-3-sites}
	-\frac{J_\textrm{d}}{2} \big( &2 \vert 2, 0, 1 \rangle \langle 1, 2, 0 \vert + \sqrt{2} \vert 0, 2, 0 \rangle \langle 1, 0, 1 \vert \nonumber \\
	&+ 2 \vert 0, 2, 1 \rangle \langle 1, 0, 2 \vert + 2 \sqrt{2} \vert 1, 2, 1 \rangle \langle 2, 0, 2 \vert + \textrm{h.c.} \big) . 
\end{align}
Observe that the off-diagonal matrix element is equal to $-J_\textrm{d} (\sqrt{2})^{n_{j-1} + n_j + n_{j+1} - 3}$, which is a function of the total number of bosons on $3$ sites.

It turns out that the dipolar Hamiltonian and its dynamics is more conveniently described by ``dipoles'' instead of the bare bosonic particles. These dipoles are naturally defined on the links of the lattice, with the dipole number operator at link $j + 1/2$ defined as a cumulative sum over boson number operators: 
\begin{equation}
	\label{eq:dipole-config-from-boson-occ}
	n_{j + \frac{1}{2}}^\textrm{d} \equiv - \sum_{k \le j} (n_k - \overline{n}),
\end{equation}
where $\overline{n} = 1$ is the average atom number per site and we set $n_{\frac{1}{2}}^{\textrm{d}} \equiv 0$ (we will always work with open boundary conditions). Note that while we refer to this as a number operator, $n_{j  + 1/2}^{\textrm{d}}$ has both positive \textit{and} negative integer eigenvalues. 

Using the dipole configurations, the total dipole number $\tilde{D}$ is given by a sum of dipole occupations on every link
\begin{equation}
	\tilde{D} = \sum_j n_{j + \frac{1}{2} }^\textrm{d}.
\end{equation}
Note that $\tilde{D}$ is related to the total dipole moment $D$ via $\tilde{D} = \overline{n} N(N + 1) / 2 - D$, where $N = \sum_j n_j$ is the total boson number. Also, note that the boson number operator may be written as a discrete derivative of dipole number operators:
\begin{equation}
	\label{eq:boson-occ-from-dipole-config}
	n_j = \overline{n} - (n_{j + \frac{1}{2}}^{\rm d} - n_{j - \frac{1}{2}}^{\rm d}). 
\end{equation} 
It will be useful to use notation in which both the boson occupations and the dipole configurations are shown simultaneously. We will thus use notation in which the dipole occupation numbers are written in red, right below the boson occupation numbers. For example, for a product state of bosons with boson numbers $\vert 1, 1, 2, 0, 1, 1 \rangle$---which has the corresponding dipole configurations $0, 0, 0, -1, 0, 0, 0$---we write 
\begin{equation}
\mathop{\vert 1,1,2,0,1,1 \rangle}_{\textcolor{red}{0, \,\, 0, \,\, 0, -1, \,\, 0, \,\, 0, \,\, 0}}.
\end{equation}
Interchanging the particle and hole in this configuration changes the sign of the corresponding dipole charge: 
\begin{equation}
	\mathop{\vert 1,1,0,2,1,1 \rangle}_{\textcolor{red}{0, \,\, 0, \,\, 0, +1, \, 0, \,\, 0, \,\, 0}}.
\end{equation}

We now analyze the implications of the boson occupation truncation at each site, $n_j \le 2$. Moreover, we will construct an effective Hamiltonian in terms of dipoles as elementary particles. It turns out that the effective Hamiltonian is an interacting model of dipoles with dipolar hopping terms, with blockades which we describe below. First of all, when there is no dipolar excitation, the dipole hopping operator acts trivially:
\begin{equation}
	{T}_{\rm d} \mathop{\vert 1, 1, 1 \rangle}_{\textcolor{red}{0, \,\, 0, \,\, 0, \,\, 0}} \approx 0 ,
\end{equation}
where we used $\approx$ to indicate that we are working in the restricted Hilbert space $n_j \le 2$. That dipoles are not created from the `vacuum' (here the state $\vert 1 \rangle^L$) by the dynamics will be important in what follows, and is due to the fact that the terms in $H_D$ involve hopping only of nearest-neighbor dipoles (including $4$-site terms as in Ref.~\onlinecite{2023Lake_DC} would spoil this property). 

Secondly, when there is a single dipolar excitation, ${T}_{\rm d}$ hops the dipolar excitation to its nearby sites:
\begin{align}
&{T}_{\rm d} \mathop{\vert 1,1,2,0,1,1 \rangle}_{\textcolor{red}{0, \,\, 0, \,\, 0,  -1, \,\, 0, \,\, 0, \,\, 0}} \approx -J_{\rm d} \big( \mathop{\vert 1, 2, 0, 1, 1, 1 \rangle}_{\textcolor{red}{0, \, 0, -1, \, 0, \,\, 0, \,\, 0, \, 0}} + \mathop{\vert 1, 1, 1, 2, 0, 1 \rangle}_{\textcolor{red}{0, \,\, 0, \,\, 0,  \,\, 0, -1, \, 0, \, 0}} \big) \nonumber \\
&{T}_{\rm d} \mathop{\vert 1,1,0,2,1,1 \rangle}_{\textcolor{red}{0, \,\, 0, \,\, 0, \,\, 1, \,\, 0, \,\, 0, \,\, 0}} \approx -J_{\rm d} \big( \mathop{\vert 1, 0, 2, 1, 1, 1 \rangle}_{\textcolor{red}{0, \,\, 0, \,\, 1, \,\, 0, \,\, 0, \,\, 0, \,\, 0}} + \mathop{\vert 1, 1, 1, 0, 2, 1 \rangle}_{\textcolor{red}{0, \,\, 0, \,\, 0, \,\, 0, \,\, 1, \,\, 0, \,\, 0}} \big)
\end{align}
Thus, in the truncated Hilbert space, isolated dipoles simply disperse freely across the system.

Things become more interesting when interactions between dipolar excitations are considered. As we have seen in Eq.~\eqref{eq:dipole-hopping-3-sites}, the hopping coefficient will be sensitive to the nearby dipole charges. When two dipolar excitations with the same dipole charges are located at next-nearest-neighboring links, the dipole hopping term still hops the dipole excitations to their neighboring links, but the hopping amplitudes now change according to Eq.~\eqref{eq:dipole-hopping-3-sites}. For two $-1$ charge dipoles, 
\begin{align}
	&{T}_{\rm d} \mathop{\vert 1,2,0,2,0,1 \rangle}_{\textcolor{red}{0, \, 0, -1, \, 0, -1, \, 0, \, 0}} \approx -J_{\rm d} \big( \mathop{\vert 2,0,1,2,0,1 \rangle}_{\textcolor{red}{0, -1, \, 0, \, 0, -1, \, 0, \, 0}} + \mathop{\vert 1,2,0,1,2,0 \rangle}_{\textcolor{red}{0, \, 0, -1, \,0, \, 0, -1, \, 0}}  \big) \nonumber \\ 
	& \qquad - J_{\rm d} \sqrt{2} \mathop{\vert 1, 1, 2, 1, 0, 1 \rangle}_{\textcolor{red}{0, \, 0, \, 0, -1, -1, \, 0, \, 0}} - \frac{J_{\rm d}}{\sqrt{2}} \mathop{\vert 1, 2, 1, 0, 1,1 \rangle}_{\textcolor{red}{0, \, 0, -1, -1, \,0, \, 0, \, 0}} .
\end{align}
and similarly for $+1$ charge dipoles, 
\begin{align}
	&{T}_{\rm d} \mathop{\vert 1,0,2,0,2,1 \rangle}_{\textcolor{red}{0, \,\, 0, \,\, 1, \,\, 0, \,\, 1, \,\, 0, \,\, 0}} \approx -J_{\rm d} \big( \mathop{\vert 0,2,1,0,2,1 \rangle}_{\textcolor{red}{0, \,\, 1, \,\, 0, \,\, 0, \,\, 1, \,\, 0, \,\, 0}} + \mathop{\vert 1,0,2,1,0,2 \rangle}_{\textcolor{red}{0, \,\, 0, \,\, 1, \,\, 0, \,\, 0, \,\, 1, \,\, 0}}  \big) \nonumber \\ 
	& \qquad - \frac{J_{\rm d}}{\sqrt{2}} \mathop{\vert 1, 1, 0, 1, 2, 1 \rangle}_{\textcolor{red}{0, \,\, 0, \,\, 0, \,\, 1, \,\, 1, \,\, 0, \,\, 0}} - J_{\rm d} \sqrt{2} \mathop{\vert 1, 0, 1, 2, 1,1 \rangle}_{\textcolor{red}{0, \,\, 0, \,\, 1, \,\, 1, \,\, 0, \,\, 0, \,\, 0}} 
\end{align}
And when two dipolar excitations are right next to each other, we have
\begin{align}
&{T}_{\rm d} \mathop{\vert 1, 2, 1, 0, 1 \rangle}_{\textcolor{red}{0, \, 0, -1, -1, \, 0, \, 0}} \approx - \sqrt{2} J_{\rm d} \mathop{\vert 2, 0, 2, 0, 1 \rangle}_{\textcolor{red}{0, \, -1, 0, -1, \, 0, \, 0}} - \frac{J_{\rm d}}{\sqrt{2}} \mathop{\vert 1, 2, 0, 2, 0 \rangle}_{\textcolor{red}{0, \, 0, -1, \, 0, -1, \, 0}} \nonumber \\
&{T}_{\rm d} \mathop{\vert 1, 0, 1, 2, 1 \rangle}_{\textcolor{red}{0, \,\, 0, \,\, 1, \,\, 1, \,\, 0, \,\, 0}} \approx - \frac{J_{\rm d}}{\sqrt{2}} \mathop{\vert 0, 2, 0, 2, 1 \rangle}_{\textcolor{red}{0, \,\, 1, \,\, 0, \,\, 1, \,\, 0, \,\, 0}} - \sqrt{2} J_{\rm d} \mathop{\vert 1, 0, 2, 0, 2, \rangle}_{\textcolor{red}{0, \,\, 0, \,\, 1, \,\, 0, \,\, 1, \,\, 0}} . 
\end{align}

When two dipolar excitations with \textit{opposite} dipole charges are nearby to each other, the constraint on the maximum number of bosons per site means that they effectively repel one another:
\begin{align}
	&{T}_{\rm d} \mathop{\vert 1,2,0,0,2,1 \rangle}_{\textcolor{red}{0, \,\, 0, -1, \, 0, \, 1, \, 0, \, 0}} \approx -J_{\rm d} \big( \mathop{\vert 2,0,1,0,2,1 \rangle}_{\textcolor{red}{0, -1, \, 0, \, 0, \, 1, \,\, 0, \,\, 0}} + \mathop{\vert 1,2,0,1,0,2 \rangle}_{\textcolor{red}{0, \, 0, -1, \, 0, \,\, 0, \, 1, \, 0}}  \big) 
\end{align}
and 
\begin{equation}
	{T}_{\rm d} \mathop{\vert 1,0,2,2,0,1 \rangle}_{\textcolor{red}{0, \,\, 0, \,\, 1, \,\, 0, -1, \, 0, \, 0}} \approx -J_{\rm d} \big( \mathop{\vert 0,2,1,2,0,1 \rangle}_{\textcolor{red}{0, \,\, 1, \,\, 0, \,\, 0, -1, \, 0, \, 0}} + \mathop{\vert 1,0,2,1,2,0 \rangle}_{\textcolor{red}{0, \,\, 0, \,\, 1, \,\, 0, \,\, 0, -1, \, 0}}  \big) . 
\end{equation}

This effective repulsion arises because configurations containing opposite-signed dipoles on adjacent links are not part of the effective Hilbert space accessible by the dynamics. Interestingly, this effect is asymmetric: states containing one $\pm 1$ dipole immediately to the left of another $\pm 1$ dipole are not allowed, but the reason is different for different choices of sign. Configurations with one $+1$ dipole immediately to the left of another $-1$ dipole are forbidden due to energy penalty coming from a large on-site Hubbard interaction, while those with one $-1$ dipole to the left of another $+1$ dipole are not possible even in principle. 

To understand what happens when the $+1$ dipole is on the left, consider the following boson configurations
\begin{equation}
	\mathop{\vert 1, 0, 3, 0, 1 \rangle}_{\textcolor{red}{0, \, 0, \, 1, -1, \, 0, \, 0}} \quad \textrm{and} \quad \mathop{\vert 1, 0, 2, 2, 0 \rangle}_{\textcolor{red}{0, \, 0, \, 1, \, 0, -1, \, 0}} . 
\end{equation}
Compared to the latter, the former has a relative energy difference $+U_\textrm{d}$. If we energetically relax the strict constraint on boson occupation numbers, then when the dipole density is low, the above energetic penalty of  $1, -1$ configuration can be captured by the following interaction term:
\begin{equation}
	U_\textrm{d} \sum_l n_{l}^{+} n_{l + 1}^{-} , 
\end{equation}
where $n_l^{+}$ ($n_l^{-}$) is the number of positively (negatively) charged dipole at link $l$, i.e., $n_l^{\textrm{d}} = n_l^{+} - n_l^{-}$. On the other hand, to realize $-1, 1$ dipole configuration, we observe that 
\begin{equation}
	\mathop{\vert 2, -1, 0 \rangle}_{\textcolor{red}{0, \, -1, \,\, 1, \,\, 0}} ,
\end{equation}
which is not allowed since $n_j = -1$ is physically impossible. Thus the dipole interactions are strongly asymmetric in space: $+$ dipoles can appear to the left of $-$ dipoles, but not the other way around.

The above observations motivate us to write down an effective model in which the dipoles are elementary excitations, which will be operative in the regime where the dipole number on each link has absolute value at most $1$. To this end, we introduce a two-species model with the corresponding creation operator $d_{l, \sigma}^\dagger$ for each link $l$ where $\sigma = +$ corresponds to positively charged dipole and $\sigma = -$ corresponds to negatively charged dipole. The total number of positive dipoles $\sum_l n_{l}^{+} = \sum_l d_{l, +}^\dagger d_{l, +}$ and negative dipoles $\sum_l n_{l}^{-} = \sum_l d_{l, -}^\dagger d_{l, -}$ are individually conserved and we impose the hardcore constraint $n_{l}^{+} + n_{l}^{-} \le 1$ at each link $l$. \footnote{One can consider an equivalent spin-$1$ chain with the conserved total $S_z$ corresponding to the total dipole moment.} Note that the dipole configuration at each link Eq.~\eqref{eq:dipole-config-from-boson-occ} becomes $n_l^{\rm d} = n_l^+ - n_l^-$ and the total dipole moment becomes $\tilde{D} = \sum_l (n_l^+ - n_l^-)$. Using the dictionary between boson occupations and dipole occupations Eqs.~\eqref{eq:dipole-config-from-boson-occ} and~\eqref{eq:boson-occ-from-dipole-config} with $\overline{n} = 1$, the on-site Hubbard term becomes
\begin{align}
	\frac{U_{\rm d}}{2} \sum_j n_j (n_j - 1) &= \frac{U_{\rm d}}{2} \sum_l \Big\{ -\big( n_{l+1}^{\rm d} - n_{l}^{\rm d} \big) + \big(n_{l+1}^{\rm d})^2 \nonumber \\
	& \qquad \qquad + \big(n_{l}^{\rm d})^2 - 2 n_{l}^{\rm d} n_{l+1}^{\rm d} \Big\} \nonumber \\ 
	&= U_{\rm d} N_{\rm d} - U_{\rm d} \sum_l n_l^{\rm d} n_{l+1}^{\rm d} ,  
\end{align}
where the summation is over link $l$ and $N_\textrm{d} = \sum_l (n_{l}^{+} + n_{l}^{-})$ is the total number of dipoles. $N_\textrm{d}$ is a conserved quantity and thus giving only a constant term. Here, we also used the hardcore constraint implying that $(n_l^\textrm{d})^2$ is either $0$ or $1$ and thus equals $n_{l}^{+} + n_{l}^{-}$. The nearest-neighbor interaction term becomes
\begin{align}
	&V_{\rm d} \sum_j n_j n_{j+1} = V_{\rm d} \sum_l \Big( 1 + (n_{l+1}^{\rm d} - n_{l}^{\rm d}) (n_{l+2}^{\rm d} - n_{l+1}^{\rm d}) \Big) \nonumber \\ 
	&= 2 V_{\rm d} \sum_j n_l^{\rm d} n_{l+1}^{\rm d} - V_{\rm d} \sum_l n_l^{\rm d} n_{l+2}^{\rm d} + \textrm{(const)} . 
\end{align}
Finally, the dipolar hopping terms become constrained hopping terms as per Eq.~\eqref{eq:dipole-hopping-3-sites}: 
\begin{align}
	&-J_d \sum_j \big( b_{j-1}^\dagger (b_j)^2 b_{j+1}^\dagger + \textrm{h.c.} \big) \nonumber \\
	&= -J_d \sum_{l, \sigma = \pm} (\sqrt{2})^{-n_{l+2}^{d} + n_{l-1}^{d}} \big( d_{l+1, \sigma}^\dagger d_{l, \sigma} + \textrm{h.c.} \big) . 
\end{align}

In sum, when the dipole density is low, the effective Hamiltonian becomes
\begin{align}
	\label{eq:dipole-effective-ham3}
	H_\textrm{d}^\textrm{eff} = &- \sum_{l, \sigma = \pm} J_{\textrm{d}, \sigma} (\sqrt{2})^{- n_{l+2}^{\textrm{d}} + n_{l-1}^{\textrm{d}}} \big( d_{l+1, \sigma}^\dagger d_{l, \sigma} + \textrm{h.c.} \big) \nonumber \\
	&- V_\textrm{d} \sum_l n_{l}^{\textrm{d}} n_{l+2}^{\textrm{d}}  - (U_\textrm{d} - 2 V_\textrm{d}) \sum_l n_l^{\textrm{d}} n_{l+1}^{\textrm{d}} \nonumber \\
	&+ U^\infty \sum_l \Big( n_{l}^{+} (n_{l}^{+} - 1) + n_{l}^{-} (n_{l}^{-} - 1) + n_{l}^{-} n_{l+1}^{+} \Big) , 
\end{align}
where $U^\infty$ is a very large coupling constant imposing the hardcore constraint at each link, $n_{l}^{+} + n_{l}^{-} \le 1$, as well the infinite repulsion for physically impossible dipole configurations $-1, 1$. We also let the dipolar hopping strength $J_{\textrm{d}, \sigma}$ depend on the dipole type $\sigma$, in order to accommodate both the higher order corrections in $\lambda$ and the phenomenology from experiments. In our case, we are in the regime where $|J_{\textrm{d}, \sigma}|, |V_\textrm{d}| \ll U_\textrm{d} \ll U^\infty$.

To better understand our effective model Eq.~\eqref{eq:dipole-effective-ham3} in the context of dipole quantum walks, we consider various static properties in the following.

\subsection{Two-dipole spectrum}

To better understand two dipole quantum walks, we study two-dipole spectra of the effective dipole Hamiltonian Eq.~\eqref{eq:dipole-effective-ham3}. We discuss how bound states appear because of strong interaction and show how our initial dipole quantum walk states spread over the scattering states and thus undergo non-trivial dynamics under time evolution. We consider both the case of two identically charged dipoles and the case of two oppositely charged dipoles.

\begin{figure*}
	\centering
	\includegraphics[width=0.9\textwidth]{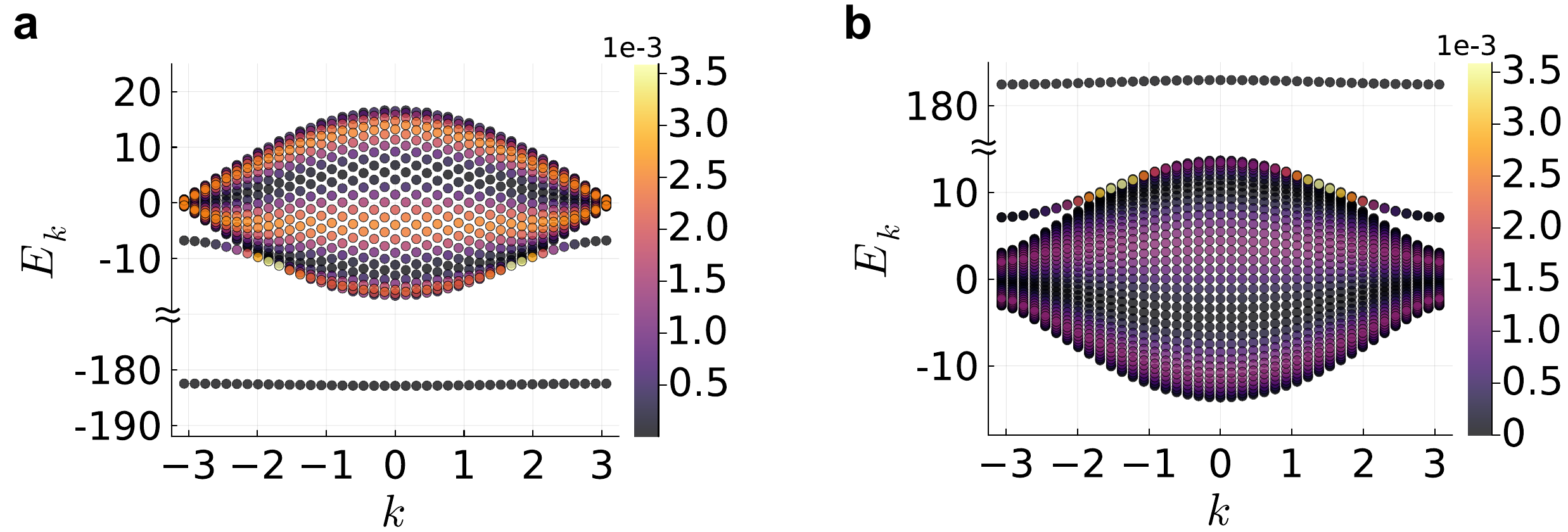}
	\caption{\textbf{Two-dipole spectrum.} Spectra of two \textbf{a,} positvely charged dipoles Eq.~\eqref{eq:dipole-pp-eigeneq} and \textbf{b,} oppositively charged dipole Eq.~\eqref{eq:dipole-pm-eigeneq} as a function of quasi-momentum $k$. The overlap squared with the eigenstates and the dipolar quantum walk initial states are also indicated. We used the system size $N=41$ and coupling constants $(J_{\textrm{d}, +}, J_{\textrm{d}, -}, V_\textrm{d}, U_\textrm{d}) = (4.18 \textrm{ Hz}, 2.67 \textrm{ Hz}, 6.79 \textrm{ Hz}, 196.0 \textrm{ Hz})$. Each spectrum consists of three bands with two being bound states and one being dispersive scattering states. Two bound states bands have energy $\tilde{U}_\textrm{d} = U_\textrm{d} - 2 V_\textrm{d}$ and $V_\textrm{d}$ away from $0$ energy. While the bound states band with energy separation $\approx \pm \tilde{U}_\textrm{d}$ has no overlap with our quantum walk initial states, the bound states band with energy separation $\approx V_\textrm{d}$ has overlap with our initial states, which is larger in case \textbf{a}.} 
	\label{FigureS_Two_Dipole_Specs}
\end{figure*}

\subsubsection{1. Two identically charged dipoles}

Here, we consider the spectrum of two positively charged dipoles of our effective dipolar Hamiltonian Eq.~\eqref{eq:dipole-effective-ham3}. The Hilbert space of two positively charged hardcore dipoles is given by $\mathcal{H}_{+, +} = \{ \vert l_1, l_2 \rangle = d_{l_1, +}^\dagger d_{l_2, +}^\dagger \vert \textrm{d-vac} \rangle : l_1 < l_2 \}$, where $\vert \textrm{d-vac} \rangle$ is the dipole vacuum state (which is equal to $\vert 1, 1, \ldots, 1 \rangle$ in terms of the boson occupation). Following the standard analysis~\cite{PhysRevA.90.062301}, we define $C_{l_1, l_2} \equiv \langle \textrm{d-vac} \vert d_{l_2, +} d_{l_1, +} \vert \Psi \rangle$ for an eigenstate $\vert \Psi \rangle$. We use periodic boundary conditions (PBC) and the ansatz solution $C_{l_1, l_2} = e^{i k \frac{l_1+l_2}{2}} \phi(l_2 - l_1)$, which decomposes into a center-of-mass plane wave part and a relative-coordinate part. Here, we impose the hardcore constraint $\phi (0) = 0$, bosonic statistics $\phi(-r) = \phi(r)$, and the PBC $\phi(L + r) = e^{i \frac{k}{2} L} \phi(r) = (-1)^\alpha \phi(r)$ with $k = \frac{2 \pi \alpha}{L}$, $\alpha \in \{1, 2, \ldots, L\}$, and assume $L$ is odd. The energy eigen equation $H_\textrm{d}^\textrm{eff} \vert \Psi \rangle  = E \vert \Psi \rangle$ translates into 
\begin{align}
	E C_{l_1, l_2} = & -J_{\textrm{d}, +} \Big( \sqrt{2}^{- \delta_{l_1+1, l_2} + \delta_{l_1-2, l_ 2}} C_{l_1 - 1, l_2} \nonumber \\ 
	& \qquad + \sqrt{2}^{- \delta_{l_1, l_2+1} + \delta_{l_1, l_2 - 2}} C_{l_1, l_2 - 1} \nonumber \\ 
	& \qquad + \sqrt{2}^{- \delta_{l_1+2, l_2} + \delta_{l_1-1, l_2}} C_{l_1+1, l_2} \nonumber \\ 
	& \qquad + \sqrt{2}^{- \delta_{l_1, l_2+2} + \delta_{l_1, l_2 - 1}} C_{l_1, l_2 + 1} \Big) \nonumber \\ 
	&-V_\textrm{d} \delta_{|l_1 - l_2|, 2} C_{l_1, l_2} - (U_\textrm{d} - 2V_\textrm{d}) \delta_{|l_1 - l_2|, 1} C_{l_1, l_2} . 
\end{align}
Using our ansatz, the eigen equation for each momentum $k$ reduces to 
\begin{align}
	E_k \phi (r) & = - 2 J_{\textrm{d}, +} \cos \big( k/2 \big) \Big( (1 - \delta_{r, 1} - \delta_{r, -2}) \phi(r + 1) \nonumber \\ 
	& \qquad \qquad \qquad \quad + (1 -\delta_{r, -1} - \delta_{r, 2}) \phi(r - 1) \Big) \nonumber \\ 
	& - J_{\textrm{d}, +} \Big( \frac{e^{-i \frac{k}{2}}}{\sqrt{2}} + \sqrt{2} e^{i \frac{k}{2}} \Big) (\delta_{r, 1} \phi(r + 1) + \delta_{r, -1} \phi(r - 1)) \nonumber \\ 
	& - J_{\textrm{d}, +} \Big( \sqrt{2} e^{-i \frac{k}{2}} + \frac{e^{i \frac{k}{2}}}{\sqrt{2}} \Big) (\delta_{r, -2} \phi(r + 1) + \delta_{r, 2} \phi(r - 1)) \nonumber \\ 
	& -V_\textrm{d} \delta_{|r|, 2} \phi(r) - (U_\textrm{d} - 2V_\textrm{d}) \delta_{|r|, 1} \phi(r) , 
\end{align}
where $k/2 = \frac{\pi \alpha}{L}$ with $\alpha \in \{1, 2, \ldots, L\}$. In terms of independent parameters, the energy eigen equation becomes 
\begin{equation}
	\label{eq:dipole-pp-eigeneq}
	E_k \vec{\phi} = \left( \begin{array}{ccccccc}
		-\tilde{U}_\textrm{d} & \tilde{J}_k & & & & & \\ 
		\tilde{J}_k^* & -V_\textrm{d} & J_k & & & & \\ 
		& J_k & 0 & J_k & & & \\ 
		& & J_k & 0 & J_k & & \\ 
		& & & \ddots & \ddots & \ddots & \\ 
		& & & & J_k & 0 & J_k \\ 
		& & & & & J_k & J_{k; \alpha} 
	\end{array} \right) \vec{\phi} , 
\end{equation}
where we define $\vec{\phi} = \big( \phi(1), \phi(2), \ldots, \phi(\frac{L-1}{2}) \big)^t$, $J_k = - 2 J_{\textrm{d}, +} \cos (k/2)$, $\tilde{J}_k = - J_{\textrm{d}, +} \big( \frac{e^{-i \frac{k}{2}}}{\sqrt{2}} + \sqrt{2} e^{i \frac{k}{2}} \big)$, $J_{k; \alpha} = (-1)^\alpha J_k$, and $\tilde{U}_\textrm{d} = U_\textrm{d} - 2 V_\textrm{d}$. 

In Fig.~\ref{FigureS_Two_Dipole_Specs}a, we compute the spectrum for $L=41$ and $(J_{\textrm{d}, +}, V_\textrm{d}, U_\textrm{d}) = h \times (4.18 \textrm{ Hz}, 6.79 \textrm{ Hz}, 196.0 \textrm{ Hz})$ together with the overlap squared with the dipolar quantum walk initial state. Due to strong interaction, there exists three bands, the one with scattering states band and two with bound states. In this case, three bands are well-separated with each other, indicating a strong interaction effect in our system. Our dipolar quantum walk initial state is distributed over the scattering states band so that the quantum walk is governed by the scattering states. 

\subsubsection{2. Two oppositely charged dipoles}

Here, we consider the spectrum of two oppositely charged dipoles of our effective dipolar Hamiltonian Eq.~\eqref{eq:dipole-effective-ham3}. The Hilbert space of two oppositively charged hardcore dipoles is given by $\mathcal{H}_{-, +} = \{ \vert l_1, l_2 \rangle = d_{l_1, -}^\dagger d_{l_2, +}^\dagger \vert \textrm{d-vac} \rangle : l_2 \ne l_1,  l_1 + 1 \}$, where we consider the hardcore constriant. We define $D_{l_1, l_2} \equiv \langle \textrm{d-vac} \vert d_{l_2, +} d_{l_1, -} \vert \Psi \rangle$ for an eigenstate $\vert \Psi \rangle$. Using the periodic boundary condition, we set the ansatz solution $D_{l_1, l_2} = e^{i k \frac{l_1+l_2}{2}} \varphi(l_2 - l_1)$, where $k = \frac{2 \pi \alpha}{L}$, $\alpha \in \{1, 2, \ldots, L\}$. Here, we impose the hardcore constraint $\varphi (0) = 0$, also impose $\varphi(1) = 0$ since $-1, 1$ is a not allowed dipole configuration, and the periodic boundary condition $\varphi(L + r) = \varphi(r)$. The eigen equation $H_\textrm{d}^\textrm{eff} \vert \Psi \rangle  = E \vert \Psi \rangle$ translates into 
\begin{align}
	E D_{l_1, l_2} = & - \Big( J_{\textrm{d}, -} \sqrt{2}^{- \delta_{l_1+1, l_2} + \delta_{l_1-2, l_2}} D_{l_1 - 1, l_2} \nonumber \\ 
	& \qquad + J_{\textrm{d}, +} \sqrt{2}^{\delta_{l_1, l_2+1} - \delta_{l_1, l_2 - 2}} D_{l_1, l_2 - 1} \nonumber \\ 
	& \qquad + J_{\textrm{d}, -} \sqrt{2}^{- \delta_{l_1+2, l_2} + \delta_{l_1 - 1, l_2}} D_{l_1+1, l_2} \nonumber \\ 
	& \qquad + J_{\textrm{d}, +} \sqrt{2}^{\delta_{l_1, l_2+2} - \delta_{l_1, l_2 - 1}} D_{l_1, l_2 + 1} \Big) \nonumber \\ 
	&+ V_\textrm{d} \delta_{|l_1 - l_2|, 2} D_{l_1, l_2} + (U_\textrm{d} - 2V_\textrm{d}) \delta_{|l_1 - l_2|, 1} D_{l_1, l_2} . 
\end{align}
Using our ansatz,
\begin{align}
	E_k \varphi (r) = & - \Big( \sqrt{2}^{- \delta_{r, 1} + \delta_{r, L-2}} \overline{J}_k \varphi(r + 1) \nonumber \\ 
	& \qquad + \sqrt{2}^{\delta_{r, L-1} - \delta_{r, 2}} \overline{J}_k^* \varphi(r - 1) \Big) \nonumber \\ 
	& +V_\textrm{d} \delta_{|r|, 2} \varphi(r) + (U_\textrm{d} - 2V_\textrm{d}) \delta_{|r|, 1} \varphi(r) , 
\end{align}
where $k/2 = \frac{\pi \alpha}{L}$ with $\alpha \in \{1, 2, \ldots, L\}$ and $\overline{J}_k = - \big( J_{d, +} e^{i k \frac{1}{2}} + J_{d, -} e^{-i k \frac{1}{2}} \big)$. In terms of independent parameters, the energy eigenequation becomes 
\begin{equation}
	\label{eq:dipole-pm-eigeneq}
	E_k \vec{\varphi} = \left( \begin{array}{ccccccc}
		V_\textrm{d} & \overline{J}_k & & & & & \\ 
		\overline{J}^*_k & 0 & \overline{J}_k & & & & \\ 
		& \overline{J}^*_k & 0 & \overline{J}_k & & & \\ 
		& & \ddots & \ddots & \ddots & & \\ 
		& & & \overline{J}^*_k & 0 & \overline{J}_k & \\ 
		& & & & \overline{J}^*_k & V_\textrm{d} & \sqrt{2} \overline{J}_k \\ 
		& & & & & \sqrt{2} \overline{J}^*_k & \tilde{U}_\textrm{d} 
	\end{array} \right) \vec{\varphi} , 
\end{equation}
where $\vec{\varphi} = \big( \varphi(2), \varphi(3), \ldots, \phi(L-1) \big)^t$ and $\tilde{U}_\textrm{d} = U_\textrm{d} - 2 V_\textrm{d}$. 

In Fig.~\ref{FigureS_Two_Dipole_Specs}b, we compute the spectrum for $L=41$ and $(J_{\textrm{d},+}, J_{\textrm{d}, -}, V_\textrm{d}, U_\textrm{d}) = h \times (4.18 \textrm{ Hz}, 2.67 \textrm{ Hz}, 6.79 \textrm{ Hz}, 196.0 \textrm{ Hz})$ together with the overlap squared with the dipolar quantum walk initial state. Due to strong interaction, there exists three bands, the one with scattering states band and two with bound states. In this case, unlike the two positively charged dipole case, one bound states band and the scattering states band overlap with each other. Moreover, our dipolar quantum walk initial state is distributed over both the scattering states and the bound states. Therefore, the quantum walk in this case is a combination of both bands.

\subsection{Dipole tunneling amplitude at non-zero $U/\Delta$}

In the previous sections, the dipole tunneling amplitude $J_{\rm d} = 2 \lambda^2 U$ was derived from a Schrieffer–Wolff transformation valid in the limit of small $\lambda = J/\Delta$ and $U/\Delta$, yielding identical tunneling rates for dipoles and antidipoles. In the experiments, however, $U/\Delta = 0.22(1)$ is only marginally small and produces a measurable asymmetry. The effective tunneling amplitudes can be obtained using second-order perturbation theory by evaluating the transition matrix elements
\begin{equation}
- \langle j_d + 1 | H^{\textrm{eff}} | j_d \rangle = - \sum_{ \substack{|\{n\}\rangle \neq \\ |j_d\rangle, |j_d+1\rangle}} \frac{ \langle j_d+1 | H_0 | \{n\} \rangle \langle \{n\} | H_0 | j_d \rangle }{ E_{ | j_d \rangle } - E_{ | \{n\} \rangle } },
\end{equation} 
where $| j_d \rangle$ denotes a single-dipole state at site $j_d$, and the sum runs over atomic Fock states $|\{n\}\rangle$ distinct from the initial and final states. The energies $E_{| j_d \rangle}$ and $E_{|\{n\}\rangle}$ are the sum of tilt and interaction energies, both diagonal in the Fock basis. Only two intermediate states contribute, obtained by moving one atom of the doublon either to the left or to the right. The resulting transition matrix element is
\begin{equation}
- \langle j_d + 1 | H^{\textrm{eff}} | j_d \rangle = - 2J^2 \left( \frac{1}{U+\Delta} - \frac{1}{\Delta} \right),
\end{equation}
which expands to $J_{\rm d}(1+U/\Delta)$ at second order in $U/\Delta$. An analogous calculation for an antidipole gives $J_{\rm d}(1-U/\Delta)$. With $U/\Delta = 0.22(1)$, the tunneling amplitudes differ by 44(2)\% between dipoles and antidipoles, in good agreement with numerical simulations. Quantitatively, the dipole tunneling amplitude is estimated as $h \times 4.1(2)$\,Hz, compared to the Schrieffer–Wolff prediction $J_{\rm d} = h \times 3.4(2)$\,Hz.

\subsection{Motzkin dynamics at unit filling}

\begin{figure}[t]
\centering
\includegraphics[width=0.5\textwidth]{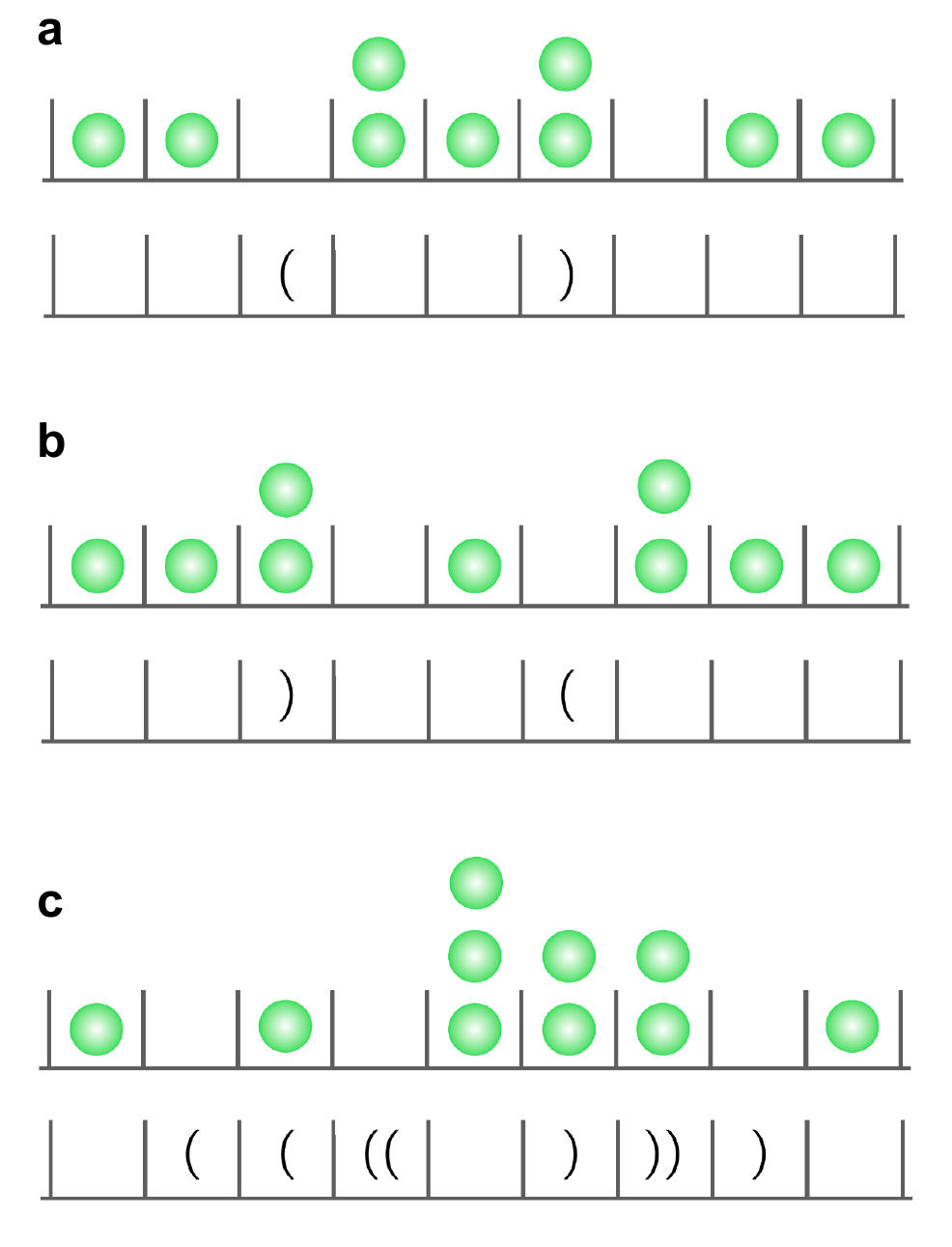}
\caption{\textbf{Examples of the mapping from bosons and dipoles (top rows) to parenthesis (bottom rows).} \textbf{a,} A dipole and an antidipole form a matched-parenthesis pair. \textbf{b,} When the positions of the dipole and antidipole are exchanged, the parenthesis pair is unmatched. Under dynamics at unit filling and with only 3-site hopping terms, this configuration is disconnected to the uniform state with one boson on each site. \textbf{c,} A more complicated example of the parenthesis mapping, showing a fully-matched state that can be accessed from the uniform state. For general configurations of bosons, more than one parenthesis may be required to live on each site.}
\label{FiguresS_Paren_Mapping}
\end{figure} 

At unit filling, the dipole-conserving Hubbard model with the three-site hopping term $b_j^\dagger (b_{j+1})^2 b_{j+2}^\dagger$ possesses a very unusual dynamical constraint at unit filling, with the pattern of dipole excitations arising under the dynamics always being mappable onto matched-parenthesis expressions. This constraint leads to Hilbert space fragmentation \cite{khemani2020localization, 2020KnapPollmann_Fragmentation}, breaking up the Hilbert space of the system into a large number of disconnected sectors, beyond just the symmetry-dictated splitting according to charge and dipole moment. Beyond serving as an experimentally-accessible example of a very exotic form of Hilbert space fragmentation \cite{khemani2020localization, 2020KnapPollmann_Fragmentation}, this is also interesting due to its connection to the constraint possessed by the famous Motzkin chain, which led to the first example of a local Hamiltonian with volume-law entanglement in its ground state \cite{2016Shor, zhang2017novel}. This mapping is explained in detail in \cite{lake2023multipole}; below we give only a brief sketch of how the mapping works. 

Let us define the field 
\[ p_x= \sum_{i=0}^x \delta \rho_i \]
where $\delta \rho_i = \rho_i - \langle \rho \rangle$ measures the departure of the boson density from its average. We will use notation where a site with $p_x = n$ hosts $n$ left parenthesis, and a site with $p_x = -n$ hosts $n$ right parenthesis. In this notation, dipoles with negative dipole moment become left parenthesis, and those with positive dipole moment become right parenthesis. Thus a state $\cdots 110212011\cdots$ is mapped to $\cdots ( \cdot \cdot )  \cdots$, while $\cdots 112010211\cdots$ becomes $\cdots)\cdot\cdot(\cdots$ (see Fig.~\ref{FiguresS_Paren_Mapping}). Note that because we can have $|p_x| > 1$, it is necessary to allow for more than one parenthesis to live on each site. 

Remarkably, the dipole-conserving Hamiltonian $H$ with hopping term $b_j^\dagger (b_{j+1})^2 b_{j+2}^\dagger$ turns out to exactly generate dynamics that preserves the {\it matchedness} of the parenthesis defined in the above fashion: it can be shown that any two configurations of parentheses with the same degree of matching can be connected by $H$, but that two configurations with different degrees of matching cannot be (the ``degree of matching'' is the number and type of parenthesis left over after removing all pairs of nested parentheses). This means that e.g., a configuration $\cdots ()\cdots$ may be dynamically connected to $\cdots (()) \cdots$ and to $\cdots ()()\cdots$, but not to $\cdots)( \cdots$ or $\cdots )()($ (the latter two of which may be connected to one another). 

To see this at a schematic level, consider what happens when the hopping term in $H$ acts on the uniform state $\cdots 11111\cdots$, producing the state $\cdots 10301\cdots$. This is mapped onto $\cdots ()\cdots$, a fully matched expression. We claim that this state can never be mapped to $\cdots )( \cdots$, even though the latter state has the same charge and dipole moment as $\cdots () \cdots$. Indeed, $\cdots )(\cdots$ corresponds to $\cdots 120021\cdots$, and since only the three-site hopping term is at our disposal, the void of two empty sites can never be crossed (see also \cite{khemani2020localization}). Thus while an antidipole to the {\it left} of a dipole can annihilate with it, if the antidipole is on the dipole's {\it right}, the two can never annihilate. This shows that under this dynamics, dipoles interact with one another in a very exotic reflection symmetry-breaking way. Further theoretical work will be required to explore the consequences this constraint has for the dynamics seen in experiment. 

For this dynamics to be realized, we require that the Hamiltonian does not contain higher-order terms like $b_j^\dagger b_{j+1} b_{j+2} b_{j+3}^\dagger$; in our setting this is a reasonable assumption at early times, since such terms are only generated at higher orders in perturbation theory. However, in order for the dynamics to be most interesting, the system also needs to be able to create and destroy nested parenthesis pairs. Starting from the uniform state, this is only possible if the system can (at least transiently) access states with three particles per site, since creating a matched parenthesis pair from $\cdots1111\cdots$ requires passing through the state $\cdots 10301\cdots$. In the regime of our experiment, where the (comparatively) large Hubbard $U$ effectively results in a prethermal conservation of doublon (or triplon) number, parenthesis creation and annihilation is suppressed, and hence different parameter ranges will need to be accessed in future work to fully explore the physics of this dynamical constraint. 

\end{document}